\documentclass[12pt]{article}
\usepackage{graphicx} % Required for inserting images
\usepackage{svg}
\usepackage{xcolor}
\definecolor{amber}{RGB}{255, 174, 55}
\usepackage[citestyle=nature]{biblatex}
\usepackage{titlesec}
\usepackage{makecell}
\usepackage[dvipsnames]{xcolor}
\usepackage{makecell} % add this to your preamble
\usepackage{booktabs} % for \addlinespace (better spacing control)
\usepackage{caption}
\usepackage{authblk}

\title{\vspace{-2em}Quantum Computing – Strategic Recommendations for the Industry}
\author[1]{Marvin Erdmann\thanks{Corresponding author: marvin.erdmann@bmw.de}}
\author[1]{Lukas Karch}
\author[2]{Abhishek Awasthi}
\author[2]{Caitlin Isobel Jones}
\author[3]{Pallavi Bhardwaj}
\author[3]{Florian Krellner}
\author[4]{Jonas Stein}
\author[4]{Claudia Linnhoff-Popien}
\author[5]{Nico Kraus}
\author[6,7]{Peter Eder}
\author[6]{Sarah Braun}
\author[6]{Tong Liu}
\affil[1]{BMW Group, Munich, Germany}
\affil[2]{BASF Digital Solutions GmbH, Ludwigshafen, Germany}
\affil[3]{SAP SE, Walldorf, Germany}
\affil[4]{LMU Munich, Munich, Germany}
\affil[5]{Aqarios GmbH, Munich, Germany}
\affil[6]{Siemens AG, Garching, Germany}
\affil[7]{Technical University Munich, Munich, Germany}
\date{}

\titleformat{\paragraph}
{\normalfont\normalsize\bfseries}{\theparagraph}{1em}{}
\titlespacing*{\paragraph}{0pt}{3.25ex plus 1ex minus .2ex}{1.5ex plus .2ex}

\newif\ifdraft
\drafttrue 
\ifdraft
\newcommand{\note}[1]{\textcolor{red}{***General: #1}}
\newcommand{\menote}[1]{ {\textcolor{blue} { ***Marvin: #1 }}}
\newcommand{\lknote}[1]{ {\textcolor{teal} { ***Lukas: #1 }}}
\newcommand{\aanote}[1]{ {\textcolor{magenta} { ***Abhishek: #1 }}}
\newcommand{\nknote}[1]{ {\textcolor{green!30!black} { ***Nico: #1 }}}

\newcommand{\jsnote}[1]{ {\textcolor{orange} { ***Jonas: #1 }}}

\else
\newcommand{\note}[1]{}
\newcommand{\menote}[1]{}
\newcommand{\lknote}[1]{}
\newcommand{\aanote}[1]{}
\newcommand{\nknote}[1]{}
\newcommand{\nkswnote}[1]{}
\newcommand{\jsnote}[1]{}

\newcommand{\linebreakand}{%
      \end{@IEEEauthorhalign}
      \hfill\mbox{}\par
      \mbox{}\hfill\begin{@IEEEauthorhalign}
    }

\fi
\begin{document}

\maketitle
\vspace{-2em}
\begin{abstract}
This whitepaper surveys the current landscape and short- to mid-term prospects for quantum-enabled optimization and machine learning use cases in industrial settings. Grounded in the QCHALLenge program, it synthesizes hardware trajectories from different quantum architectures and providers, and assesses their maturity and potential for real-world use cases under a standardized traffic-light evaluation framework.
We provide a concise summary of relevant hardware roadmaps, distinguishing superconducting and ion-trap technologies, their current states, modalities, and projected scaling trajectories. The core of the presented work are the use case evaluations in the domains of optimization problems and machine learning applications. For the conducted experiments, we apply a consistent set of evaluation criteria (model formulation, scalability, solution quality, runtime, and transferability) which are assessed in a shared system of three categories, ranging from optimistic (solutions produced by quantum computers are competitive with classical methods and/or a clear path to a quantum advantage is shown) to pessimistic (significant hurdles prevent practical application of quantum solutions now and potentially in the future). The resulting verdicts illuminate where quantum approaches currently offer promise, where hybrid classical-quantum strategies are most viable, and where classical methods are expected to remain superior. 
\end{abstract}
%\footnote[*]{corresponding author: marvin.erdmann@bmw.de}
\newpage
% \begin{figure}[htbp]
%     \centering
%     \includegraphics[width=\linewidth]{figures/acl_scaling.pdf}
%     \caption{The ACL scaling plot shows the number of variables as a function of the number of cars to be loaded. The quadratic model can handle quadratic constraints. All models show linear scaling.
% }
%     \label{fig:acl_scaling}
% \end{figure}

% \begin{figure}[htbp]
%     \centering
%     \includegraphics[width=\linewidth]{figures/mpl_scaling.pdf}
%     \caption{The scaling plot for MoProLog reveals a fast-increasing number of QUBO variables even for small problems. The MILP formulation is equal to the theoretical scaling. Small and large refer to short and long processing times of the jobs.}
%     \label{fig:moprolog_scaling}
% \end{figure}

% https://thequantuminsider.com/2025/05/16/quantum-computing-roadmaps-a-look-at-the-maps-and-predictions-of-major-quantum-players/

% \newpage
\section{Introduction}
%\note{Check spelling of words like QCHALLenge, use case, AE/BE words, and abbreviate only once...}
The project "Quantum-Classical Hybrid Optimization Algorithms for Logistics and Production Line Management" (QCHALLenge), funded by the German Federal Ministry of Research, Technology and Space (BMFTR), explores the industrial applicability of quantum computing (QC) in optimization and machine learning~\cite{qchallenge}. Led by a consortium including BASF, BMW Group, SAP SE, Siemens, and LMU Munich, the project aims to evaluate quantum-classical hybrid algorithms for real-world production and logistics challenges. Several industrial use cases are tested on different quantum algorithms, along with their implementations on noisy intermediate-scale quantum (NISQ) devices, focusing on hybrid software that mitigates hardware limitations such as low qubit counts, limited connectivity, and error susceptibility. These algorithms are benchmarked against classical solvers to test efficacy of quantum algorithms and the quantum hardware. Benchmarking is central to the project’s technical roadmap. During the course of this project, we utilize quantum hardware from IBM, IonQ, DWave~\cite{PAS,moprolog}. %\note{Consider adding references.}

Key aspects of the QCHALLenge projects are:
\begin{itemize}
    \item A top-down approach and focuses on application-layer tools to reduce complexity for non-experts.
    \item Development of a hybrid analysis and evaluation tool to assess problem complexity and performance across QC hardware.
    \item Emphasis on logistics and production due to their high industrial relevance and repeatability.
\end{itemize}

This report is part of one of the overarching goals of QCHALLenge in providing methodological recommendations for the industry, grounded in the estimated business impact of QC on selected use cases in production and logistics. This catalog also includes the QC roadmaps of several hardware vendors tested during this project, and their suitability for use cases provided by the project members. Please note, that we refer to the hardware vendors' respective technology roadmaps publicly available on the date of publication of this work.

% Who are we? 
% What did we do? 
% How do we fit into the quantum ecosystem? 
The QCHALLenge project is part of a broader quantum ecosystem and connects the following other projects, initiatives and industries.
\begin{itemize}
\item PlanQK: Commercial quantum platform focused on small and medium-sized enterprises; QCHALLenge complements it by targeting large industry~\cite{planqk}.
\item Luna: Commercial quantum solution suite provided by Aqarios, a spin-off from the Quantum Applications and Research Laboratory (QAR-Lab) at LMU Munich; They collaborated with all members of QCHALLenge and supported the research for this study~\cite{luna}.
\item QUARK: Quantum computing application benchmarking framework; QCHALLenge provides public plugin with its use cases and metrics~\cite{quark_github}.
\item QuCUN: User-oriented quantum computing platform; QCHALLenge contributes tools and demonstrators~\cite{qucun}.
\item QUTAC: Industry consortium; QCHALLenge aligns with its mission to industrialize quantum computing~\cite{QUTAC}.
\item Events \& Dissemination: QCHALLenge contributes to high-profile conferences and public events like the  Bitkom Quantum Summit and multiple QCE conferences and informs the public about these events on its own webpage~\cite{bitkom_talk1, QCE_2024, QCE_2025}.
\end{itemize}

The remainder of the paper is structured as follows: Section~\ref{sec:hardware} presents the publicly available technology roadmaps of selected quantum hardware providers with a focus on quantum architectures that were used in the experiments conducted for this study, namely superconducting devices (Section~\ref{sec:SCQC}) and ion traps (Section~\ref{sec:ion_traps}). In Section~\ref{sec:use_cases}, we first introduce the categories and the evaluation framework used to assess the performance and potential of quantum approaches in this work. Following that, we describe industry applications from the domains of optimization problems (Section~\ref{sec:opt}) and machine learning (Section~\ref{sec:QML_use_cases}), assess the capabilities of quantum solutions, and make a verdict on the potential of quantum computing for each case. We conclude with an overview (Section~\ref{sec:overview}) and outlook on the promises of quantum computing for industrial use cases based on our results and give recommendations for users of the evaluated technologies (Section~\ref{sec:outlook}).

\section{Quantum Hardware Roadmap}
\label{sec:hardware}
% \note{Short intro, including significance of hardware roadmaps for industry and research, and clarification that we are aware of the fact that qubit counts are not the only (and probably not even the best) metric to compare the scale and capabilities of quantum hardware, but certainly one of the easiest to understand. It will also be the scale we use in Section~\ref{sec:use_cases} to evaluate the potential of industrial use case for short-term quantum advantage~\cite{quantum_insider_roadmap}.
% We should also mention other types of quantum hardware, like photonic or topological quantum computers, and why we do not consider them in our studies (focus on the technologies we used for the experiments with our use cases).}

Quantum hardware roadmaps are a foundational element for both industry and research, enabling strategic planning, prioritization, and disciplined resource allocation across projects, budgets, and time horizons. A reliable roadmap translates technical capability ceilings into actionable milestones, supports risk assessment, and guides decisions on experimentation, procurement, and collaboration. By aligning anticipated hardware evolutions with concrete use cases, organizations can anticipate near-term feasibility, adjust expectations for mid- to long-term impact, and orchestrate the allocation of expertise, resources, and funding accordingly. In this sense, robust roadmaps are the basis for the entire development pipeline from research to deployment.

Qubit counts, while the most transparent and widely communicated metric offered by vendors, constitute only one dimension of quantum hardware capacity. Other critical factors are also relevant for practical performance: qubit quality, including coherence times and gate fidelities; qubit connectivity and the architectural layout that determines embeddability; error rates and the prospects for error correction; the breadth of native gate sets; and the maturity of software stacks, including compilers, embedding techniques, and ecosystem tooling~\cite{metrics2020009}. Nevertheless, for the purposes of this study, we employ qubit counts as the core metric. This choice reflects the practical advantages of qubit-count comparisons: they offer a straightforward, consistently reported benchmark that facilitates the evaluation across diverse hardware platforms and provides an intuitive basis for long-range planning and cross-site benchmarking.

This section focuses on the product roadmaps of providers that build two kinds of quantum technology. Section~\ref{sec:SCQC} surveys superconducting quantum computers, detailing their current state, the coexistence of gate-based and annealing modalities within this modality, and the prevailing scaling trajectories in terms of qubit counts. Section~\ref{sec:ion_traps} turns to ion-trap-based quantum computers, describes their physical concept, and presents the technology roadmaps of providers of such devices. Alternative quantum-related hardware families, such as neutral-atom-based~\cite{NAQC}, photonic~\cite{PsyQuantum} and topological quantum computers~\cite{Majorana}, were not used for the evaluation of the use cases in this study and are therefore not in the focus of this section.

\subsection{Superconducting Quantum Computers}
\label{sec:SCQC}

Superconducting devices represent a prominent class of quantum computing technology, distinguished by their use of superconducting materials to create qubits. These qubits operate at extremely low temperatures, where materials exhibit zero electrical resistance, allowing for the efficient manipulation of quantum states. Superconducting qubits are typically implemented using Josephson junctions, which enable the creation of non-linear inductive elements that are essential for quantum operations.

One of the key advantages of superconducting devices is their compatibility with existing semiconductor fabrication techniques, which can accelerate the development and scaling of quantum systems. This compatibility allows for the potential mass production of superconducting qubits, making them a viable option for future quantum computing applications. However, challenges remain, particularly in terms of managing the thermal environment and minimizing noise, which can adversely affect qubit performance.

Superconducting qubits are one of several ways how to realize quantum computing hardware (another are ion traps which are introduced in Section~\ref{sec:ion_traps}). Two computational methods that use superconducting qubits are quantum annealing and gate-based devices.

One of the primary distinctions between gate-based quantum computing and quantum annealers lies in their operational principles. Gate-based systems rely on a sequence of quantum gates to perform calculations. These gates manipulate the quantum states of qubits, allowing for the implementation of universal quantum algorithms. The flexibility of gate-based quantum computers enables them to tackle a broad spectrum of problems, from factoring large numbers to simulating quantum systems.

In contrast, quantum annealers are specifically designed for optimization problems. They operate by finding the lowest energy state of a system, which corresponds to the optimal solution of a given problem. Quantum annealers do not utilize a series of gates; instead, they leverage quantum tunneling and superposition to explore multiple solutions simultaneously. This makes them particularly effective for specific types of optimization tasks, but less versatile than gate-based systems for general-purpose quantum computing.

\subsubsection{Quantum Annealers}
\label{sec:annealers}
Quantum annealing is a distinct approach within the realm of quantum computing, setting itself apart from other quantum technologies such as gate-based quantum computing. In contrast to other quantum technologies that may require complex circuit designs and extensive error correction, quantum annealing offers a more direct pathway to solving specific types of optimization problems, making it particularly suited for applications where finding optimal solutions is critical.

The technology roadmap for quantum annealing has seen significant advancements, particularly through the efforts of companies like D-Wave~\cite{dwave, dwave_advantage}, which has been at the forefront of this technology. The evolution of quantum annealers has been marked by a steady increase in qubit count and improvements in hardware capabilities. This growth is crucial as the number of qubits directly impacts the complexity of problems that can be addressed. However, the scaling of qubits is not merely a linear progression; it involves overcoming significant challenges related to embedding problems into the quantum architecture and ensuring that the qubits maintain coherence long enough to perform calculations.

The roadmap for quantum annealing also emphasizes the need for enhanced error correction and noise reduction techniques. Current quantum annealers operate in the NISQ era, where qubits are prone to errors that can affect the reliability of solutions. As such, future developments will focus on improving qubit fidelity and developing more sophisticated algorithms that can effectively utilize the available qubits.

Moreover, the anticipated advancements in quantum annealing are not just about increasing qubit counts. They also involve refining the algorithms used to translate real-world problems into a quadratic unconstrained binary optimization (QUBO) model suitable for quantum processing. This transformation is often complex and can introduce inefficiencies that diminish the potential advantages of quantum solutions over classical methods.

\begin{figure}[htbp]
    \centering
    \includegraphics[width=1\textwidth]{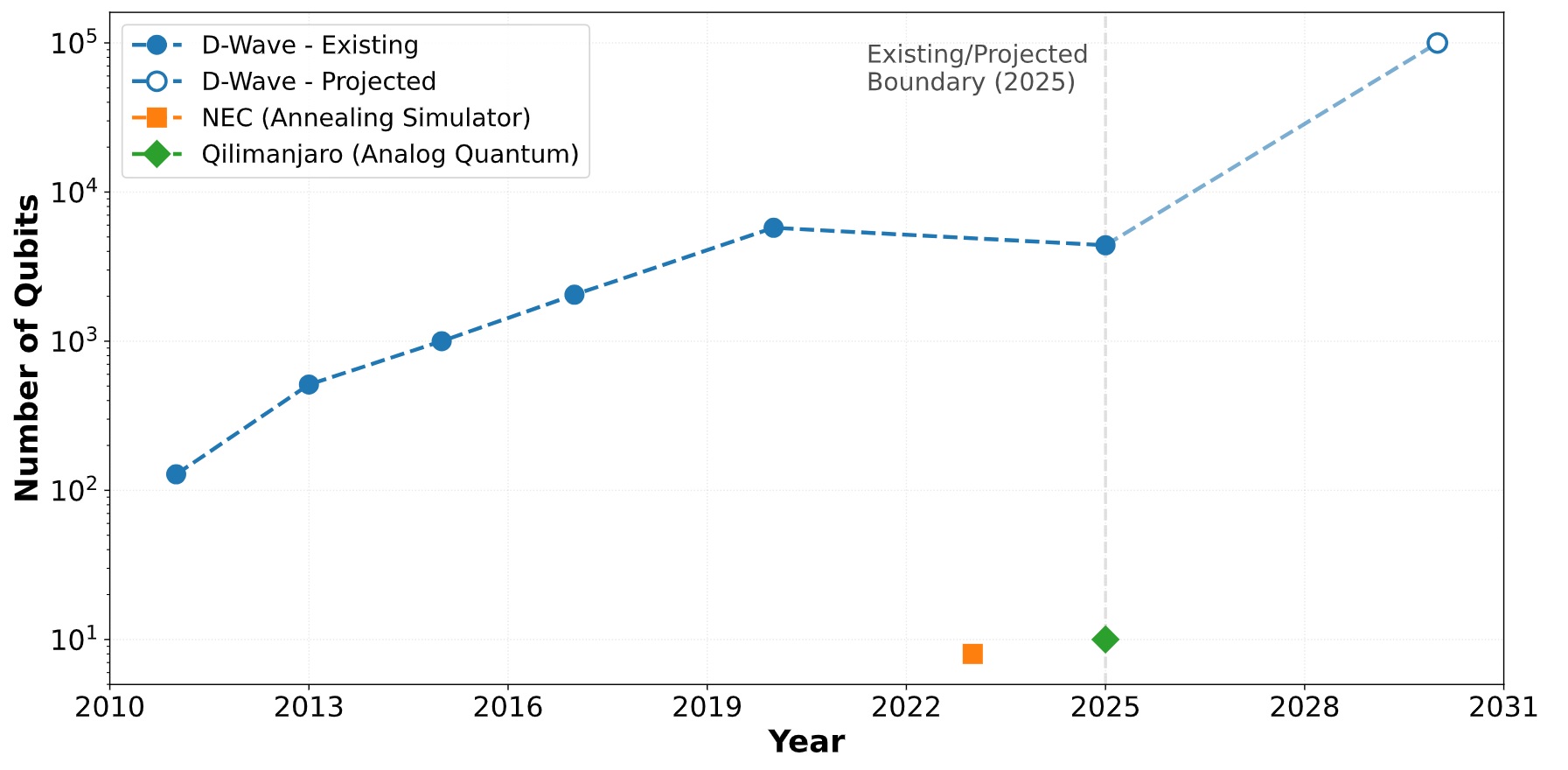}
    \caption{Historical and projected qubit count development of D-Wave's quantum annealing devices (blue) compared with qubit counts of devices manufactured by NEC (orange) and Qilimanjaro (green).}
    \label{fig:annealer_qubit_counts2}
\end{figure}

Figure~\ref{fig:annealer_qubit_counts2} depicts the historic and advertised qubit counts according to the technology roadmap of D-Wave, as well as the qubit counts of devices provided by NEC~\cite{nec} and Qilimanjaro~\cite{qilimanjaro}, two smaller companies from Japan and Spain, respectively, with a footprint in quantum annealing.

% \textcolor{red}{Where do these numbers come from, especially those for NEC and Qilimanjaro? All I can find (https://parityqc.com/a-new-quantum-annealer-by-nec, https://thequantuminsider.com/2025/01/28/eurohpc-ju-signs-procurement-contract-for-europes-first-quantum-annealer-set-to-launch-in-spain/) points towards current machines with not more than 10 physical qubits and no mention of future plans what so ever? Please provide sources or adjust the image accordingly!}

% \textcolor{red}{For D-Wave, the numbers are also questionable: Currently, their largest Advantage 2 device offers 4400 qubits (not 7440) and I cannot find any official roadmap that states intermediate steps until they 2030 when they aim for 100k physical qubits. Can you provide any source that mentions anything else?}

% \textcolor{red}{I added a figure that reflects all data points I could find sources for (Fig~\ref{fig:annealer_qubit_counts2}).  I removed the prototype versions of Advantage~2 from the new graph for comparability.}

With a history of 100+ qubit quantum annealers since 2011, D-Wave is the leading manufacturer of such devices since the very beginning of this technology's engineering. Their largest device in terms of the number of qubits is \textit{Advantage} which was made publicly available in 2020 and features 5760 physical, superconducting flux qubits~\cite{d-wave-advantage, Harris_2010}. Since then, D-Wave worked on the successor, \textit{Advantage2}, which was made fully publicly available in 2025 focusing on increasing the qubit interconnectivity and coherence, and reducing noise~\cite{d-wave-advantage2}. While currently being limited to about 4400 physical qubits, D-Wave communicated their plans to update their devices continuously, aiming for a 100,000 qubit system in 2030~\cite{d-wave-roadmap}.

Outside of D-Wave, the landscape of quantum annealing hardware pro\-viders is very scarce. NEC Corporation, Tohuku University, Japan’s National Institute of Advanced Industrial Science and Technology, and ParityQC all partnered in 2023 to build an 8-qubit quantum annealing machine, mainly used for public research~\cite{nec}. Qilimanjaro went a step beyond that when they integrated a 10-qubit adiabatic quantum processing unit into the Barcelona Supercomputing Center in 2025, making it the first high performance computing (HPC-)integrated quantum annealing device in Europe~\cite{qilimanjaro, qilimanjaro2}. However, no future development plans have been communicated by any quantum annealing hardware provider but D-Wave.

% \begin{figure}[htbp]
%     \centering
%     \includesvg[width=1\textwidth]{figures/quantum_annealing_qubit_projections.svg}
%     \caption{Historical and projected qubit count development of various quantum annealing model providers.}
%     \label{fig:annealer_qubit_counts}
% \end{figure}

In summary, the future success of quantum annealing will rely on its advancements in qubit scaling but maybe even more so in overcoming challenges like limited qubit connectivity, error correction, and algorithmic efficiency. As the technology matures, it may turn out to be the directest path to a practical quantum advantage by solving complex optimization problems across various industries. However, achieving these advancements will require continued research and development to address the inherent challenges of quantum hardware and algorithm design. The fact that with D-Wave there is only one globally relevant provider of quantum annealing hardware, makes this technology relatively vulnerable to the commercial success and stability of this player in a highly dynamic ecosystem.

\subsubsection{Gate-based Devices}
\label{sec:gate-based}
Gate-based quantum computing is a versatile approach that utilizes quantum gates to manipulate qubits, enabling the execution of a wide range of quantum algorithms. This method is characterized by its ability to perform complex computations through a series of quantum operations, making it suitable for various applications, including quantum simulations, cryptography, and optimization problems.

The roadmaps for gate-based devices from providers like IBM~\cite{ibm_roadmap} and IQM~\cite{iqm_roadmap} foresee rapid advancements, particularly in improving coherence times, gate fidelities, and the number of physical (and later also logical) qubits. As researchers continue to refine the design and fabrication processes, the performance of superconducting qubits has steadily improved, enabling more complex quantum algorithms to be executed. Additionally, the modular nature of gate-based circuits allows for the potential integration of various qubit types and classical control systems, enhancing the overall functionality of quantum processors.

\begin{figure}[htbp]
    \centering
    \includegraphics[width=\textwidth]{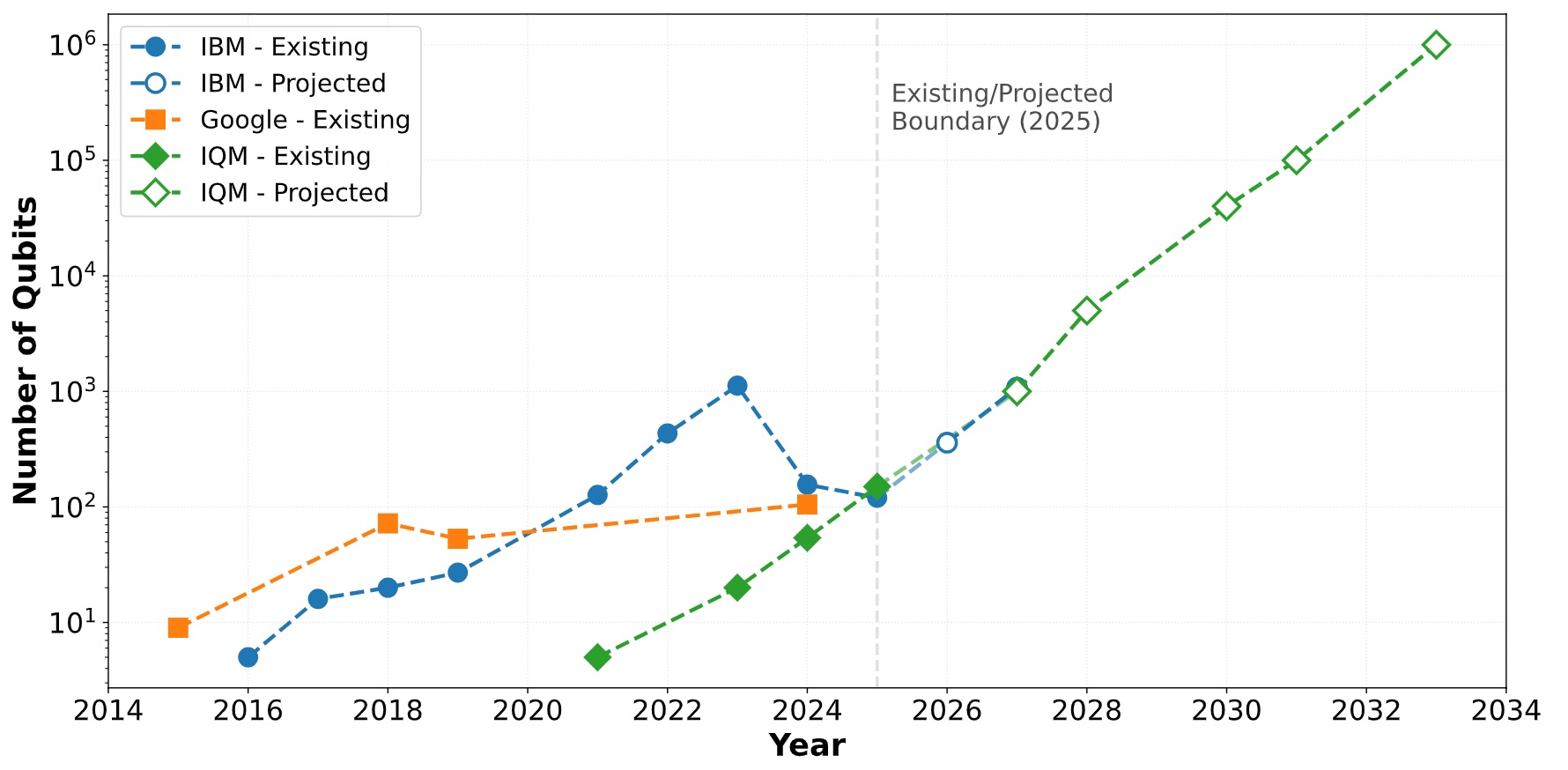}
    \caption{Historic and projected development of the physical qubit counts of gate-based quantum computers manufactured by IBM (blue), Google (orange), and IQM (green)}
    \label{fig:sc_gate_qubit_counts2}
\end{figure}

% Paragraph on roadmap and providers (IBM,IQM, Google)
The graph in Figure~\ref{fig:sc_gate_qubit_counts2} includes the historical development of gate-based quantum computers built by three of the leading providers of this technology (IBM, Google, and IQM) and the projected development according to their respective technology roadmaps.

Compared to quantum annealing, the landscape of providers for gate-based quantum computers is significantly more dense when it comes to qubit counts. While Google held the record for the most qubits in a gate-based quantum computer between 2015 and 2020~\cite{google_9qubits, google_bristlecone}, including the \textit{Sycamore} processor with which Google claimed quantum supremacy in 2019~\cite{google_sycamore}, IBM's \textit{Eagle} processor was the first gate-based quantum computing device with more than 100 qubits in 2021~\cite{ibm_eagle}. In the same year, IQM provided access to their 5-qubit \textit{Spark} devices~\cite{iqm_spark}, joining the race for leadership between manufacturers of superconducting gate-based quantum computers. In 2023, IBM built the largest such device to date: \textit{Condor} is a 1121-qubit quantum processor with which IBM announced "the beginning of the era of quantum utility"~\cite{ibm_condor}.

Since then, IBM focused on single-qubit quality improvements, qubit connectivity, and modularity, leading to the development of smaller but more powerful chips in 2024 (\textit{Heron}~\cite{ibm_heron, ibm_heron2}) and 2025 (\textit{Nighthawk}~\cite{ibm_nighthawk}). In that time frame, Google and IQM closed the gap with Google \textit{Willow} (105 qubits)~\cite{google_willow} and IQM \textit{Radiance} (150 qubits)~\cite{iqm_radiance}.

According to the hardware roadmaps of IBM~\cite{ibm_roadmap} and IQM~\cite{iqm_roadmap}, both providers plan to further scale their quantum machines. Google also published a list of milestones they plan to achieve but no concrete timeline~\cite{google_roadmap}. All of them agree on the goal to build an error-corrected, fault-tolerant quantum computer within this decade, starting to formulate their target numbers in terms of logical qubits rather than physical qubits at some point during their respective roadmaps (we only consider communicated projections for physical qubits in the figures presented in this work). This clear path towards fault-tolerance allows to estimate the date at which industry-relevant problems can be efficiently solved on quantum hardware and will be the basis on which we evaluated the use cases in Section~\ref{sec:use_cases}.

% \textcolor{red}{Sources I found and used for Fig~\ref{fig:sc_gate_qubit_counts2}: https://meetiqm.com/press-releases/iqm-quantum-computers-unveils-development-roadmap-focused-on-fault-tolerant-quantum-computing-by-2030/, https://quantumai.google/roadmap, https://www.ibm.com/quantum/blog/large-scale-ftqc}

% \begin{figure}[htbp]
%     \centering
%     \includesvg[width=\textwidth]{figures/sc_gate_qubit_projections.svg}
%     \caption{Qubit count historic and projected development of different superconducting gate model providers.}
%     \label{fig:sc_gate_qubit_counts}
% \end{figure}

In summary, superconducting devices are among the leading technologies in the quantum computing landscape, characterized by their broad spectrum of application domains and rapid advancements in performance. As research progresses, these devices are expected to play a crucial role in the realization of practical quantum utilization and a potential quantum advantage, capable of tackling complex, large-scale industry problems that are currently beyond the reach of classical computing technologies.

\subsection{Ion Traps}
\label{sec:ion_traps}
Ion trap technology is a different approach to realize qubits in the field of quantum computing compared to superconducting devices. Its distinguishing feature is its use of charged particles, or ions, as qubits. Unlike superconducting devices, which rely on solid-state systems, ion traps utilize electromagnetic fields to confine and manipulate individual ions in a vacuum. This method allows for high-fidelity quantum operations, as ions can be precisely controlled using laser beams to perform quantum gates. The inherent stability of trapped ions contributes to longer coherence times compared to many other qubit implementations, making them particularly attractive for processes that require many quantum gate operations per qubit.

One of the primary differences between ion traps and superconducting devices lies in their operational environment. Ion traps operate at room temperature, simplifying the cooling requirements compared to superconducting qubits that must be maintained at cryogenic temperatures. This characteristic can facilitate easier integration with classical control systems and may reduce the complexity of the overall quantum computing architecture.

Additionally, ion trap systems typically exhibit a high degree of connectivity between qubits, allowing for more flexible quantum circuit designs. This contrasts with superconducting devices, where qubit connectivity can be limited by the physical layout of the chip. The ability to perform entangling operations between any pair of ions in a trap enhances the potential for implementing complex quantum algorithms.

While offering some theoretical and practical advantages to other qubit technologies, historically, trapped-ion-based quantum computers also faced unique challenges, particularly in terms of scalability. While individual ion traps can achieve excellent performance, creating large-scale quantum processors with many qubits remains a significant hurdle for hardware providers like Quantinuum, IonQ, and AQT. The complexity of managing multiple ions and ensuring precise control over their interactions can complicate the scaling process. %In contrast, superconducting devices benefit from their ability to integrate many qubits on a single chip, which may provide a more straightforward path to scaling.

\begin{figure}[htbp]
    \centering
    \includegraphics[width=\textwidth]{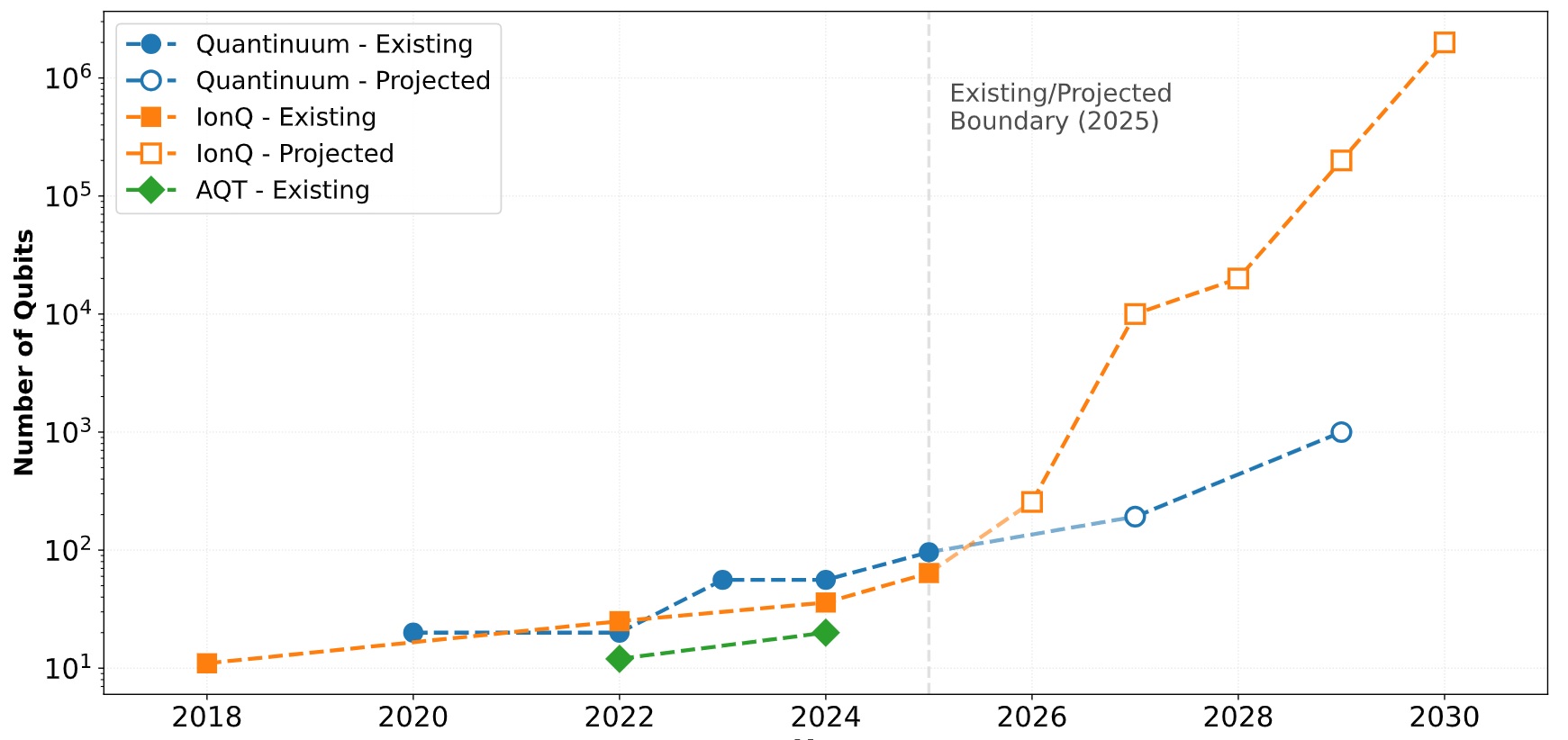}
    \caption{Historic and projected development of trapped-ion-based quantum computers manufactured by Quantinuum (blue) and IonQ (orange), compared with qubit counts of devices manufactured by AQT (green).}
    \label{fig:ion_gate_qubit_counts2}
\end{figure}

The two leading trapped-ion quantum computing manufacturers Quantinuum and IonQ advertise considerable increases in qubit counts of upcoming devices until the year 2030, while AQT does currently not provide a public hardware roadmap. Figure~\ref{fig:ion_gate_qubit_counts2} shows that, in 2025, the largest trapped-ion-based quantum computer in terms of qubits is Quantinuum Helios with 96 physical Barium-137 qubits~\cite{quantinuum_roadmap2}. IonQ advertises a device with 2~Million individual ions within five years from now~\cite{ionq_roadmap}, while Quantinuum expects a more moderate growth to ``1000s physical qubits,'' focusing on reducing the logical error rates, according to their latest technology roadmap~\cite{quantinuum_roadmap}.

% \textcolor{red}{Sources for Fig.~\ref{fig:ion_gate_qubit_counts2}: https://www.quantinuum.com/press-releases/quantinuum-unveils-accelerated-roadmap-to-achieve-universal-fault-tolerant-quantum-computing-by-2030, https://ionq.com/blog/ionqs-accelerated-roadmap-turning-quantum-ambition-into-reality}

% \begin{figure}[htbp]
%     \centering
%     \includesvg[width=\textwidth]{figures/ion_gate_qubit_projections.svg}
%     \caption{Qubit count historic and projected development of different ion-based gate model providers.}
%     \label{fig:ion_gate_qubit_counts}
% \end{figure}

In summary, ion trap technology offers a distinct approach to quantum computing, characterized by its use of trapped ions as qubits and its operational advantages in terms of coherence and connectivity. As research continues to advance in this area, ion traps are expected to contribute significantly to the development of practical quantum computers, complementing the capabilities of superconducting devices.

% \subsection{Other Types of Qubits}
% \label{sec:other_devices}

\section{Industrial Use Cases Considered}
\label{sec:use_cases}
One of the most important tasks in the QCHALLenge project is the use case identification and evaluation in the field of combinatorial optimization and machine learning with quantum computing. Each use case is assessed using both classical and QC-based approaches, with an estimation of benefits with quantum computing and further steps required in obtaining a quantum advantage.

The code basis used for the experiments conducted for the project is available as an open-source GitHub repository~\cite{qchallenge_github}. The results can also be reproduced and benchmarked with other solution methods and (quantum) hardware backends due to its implementation as a plugin of the Quantum Computing Application Benchmarking (QUARK) framework~\cite{quark_github}.

\subsection{Traffic-Light-based Evaluation}
\label{sec:traffic-light-system}
\subsubsection{Considered Aspects of Use Cases}
\label{sec:aspects}

Evaluating quantum computing through a set of well-defined, standardized aspects provides a structured and transparent way to understand its capabilities, limitations, and potential for practical application. These aspects help identify where quantum approaches already offer advantages, where significant challenges remain, and what areas require further research. By applying consistent criteria across diverse use cases, it becomes possible to make meaningful comparisons, prioritize development efforts, and understand the realistic prospects for industrial implementation. Additionally, in some instances, specific use cases may require the examination of supplementary, use-case-specific aspects. These additional evaluations are intended to highlight particular advantages, disadvantages, or unique properties of individual applications, rather than to serve as a basis for direct comparison with other use cases.
\begin{itemize}
    \item \textbf{Model and Implementation}:  This aspect evaluates how the problem is formulated and represented in the quantum computing context, including the specific algorithms, encodings, and problem structures used. It measures how well the problem fits into quantum-friendly formats like QUBO or a binary quadratic program (BQP) and how effectively the model can be implemented on quantum hardware, e.g., if slack variables are needed to implement constraints and if they hinder the performance.
    \item \textbf{Scalability}: This aspect assesses the ability of quantum computing approaches to handle increasing problem sizes. It considers the number of variables, qubits, and the computational resources needed as the problem complexity grows, and whether current or future hardware can support large-scale industrial problems.
    \item \textbf{Solution Quality}: This measures the optimality, accuracy, and stability of solutions produced by quantum algorithms compared to classical methods. It evaluates whether quantum approaches can find better, faster, or more reliable solutions in practice.
    \item \textbf{Runtime}: This aspect examines the time efficiency of quantum solutions in comparison to classical methods. It considers the total time taken for problem solving, including embedding, execution, and post-processing, and its suitability for real-time or near-real-time applications.
    \item \textbf{Transferability}: This evaluates the applicability of quantum solutions across different problem types or domains. It reflects the flexibility of the evaluated quantum methods to adapt to various use cases and the potential for broad industrial deployment.
\end{itemize}
\subsubsection{Categories, Evaluation, and Verdict}
\label{sec:categories}
The traffic-light-based system is a straightforward and intuitive method for categorizing the performance and potential of quantum computing across different aspects. By using the colors green, yellow, and red, this system quickly conveys the current state, short-term prospects, and long-term challenges associated with each aspect. Green indicates that quantum approaches are competitive or promising, yellow signifies areas where benefits are uncertain or depend on future developments, and red highlights aspects where significant hurdles prevent practical application now and potentially in the future. This visual and easily understandable framework facilitates clear communication of complex technical evaluations, helping stakeholders quickly grasp the strengths and limitations of quantum computing in various use cases. It also supports strategic decision-making by highlighting priority areas for research and development, as well as identifying aspects that require long-term investment and innovation, which is provided at the conclusion of each use case section in the \textit{verdict}.
\begin{itemize}
    \item \textbf{Green Light}: In these aspects, quantum computing (or hybrid approaches) is competitive to or better than classical methods evaluated for this use case. There is a clear path to a tangible short-term quantum advantage.
    \item \textbf{Yellow Light}: It remains to be seen if and when quantum computers can compete or provide a benefit in aspects categorized as yellow compared to classical solutions.
    \item \textbf{Red Light}: The results produced in this study clearly indicate that quantum computing cannot compete with classical methods in regards of these aspects. According to the available technology roadmaps, there is no clear path to quantum advantage within the next five to eight years.
\end{itemize}

\subsection{Optimization Use Cases}
\label{sec:opt}
The QCHALLenge project tackles a diverse set of industrial optimization use cases, each selected for its complexity, economic relevance, and potential for quantum advantage. These use cases span multiple industrial consortium partners, and are designed to benchmark quantum computing approaches against classical methods. Each use case is rigorously benchmarked on both simulated and real quantum hardware (if possible), including D-Wave, IBM and IonQ quantum computers, to assess feasibility and performance.

In the next sections, we present a short description of the use cases, along with the modeling of the problems for classical and quantum solvers, as the mathematical model of a problem is one of the most important aspects of any solving method (classical or quantum). We then present the overview of the evaluation of different aspects of the use case as described above. %on the solution quality (comparing the classical and quantum solutions) and the runtime aspects.
%Henceforth we present the feasibility of the use cases on current quantum hardware and software. We then summarize the usefulness of quantum computing for the use cases for an industrial benefit.

% The way a problem is structured determines:
% \begin{itemize}
% \item Solver compatibility: Classical solvers (e.g. mixed-integer programming, Monte Carlo methods) and quantum solvers (e.g. QAOA, quantum annealing) have different strengths. A well-modeled problem aligns with the solver’s capabilities.
% \item Computational efficiency: Poor modeling can lead to unnecessary complexity, making even powerful solvers inefficient.
% \item Solution quality: The formulation affects not just whether a solution is found, but how good that solution is.
% \end{itemize}

\subsubsection{Production Assignment and Scheduling}
\begin{figure}[htbp]
    \centering
    \includegraphics[width=\textwidth]{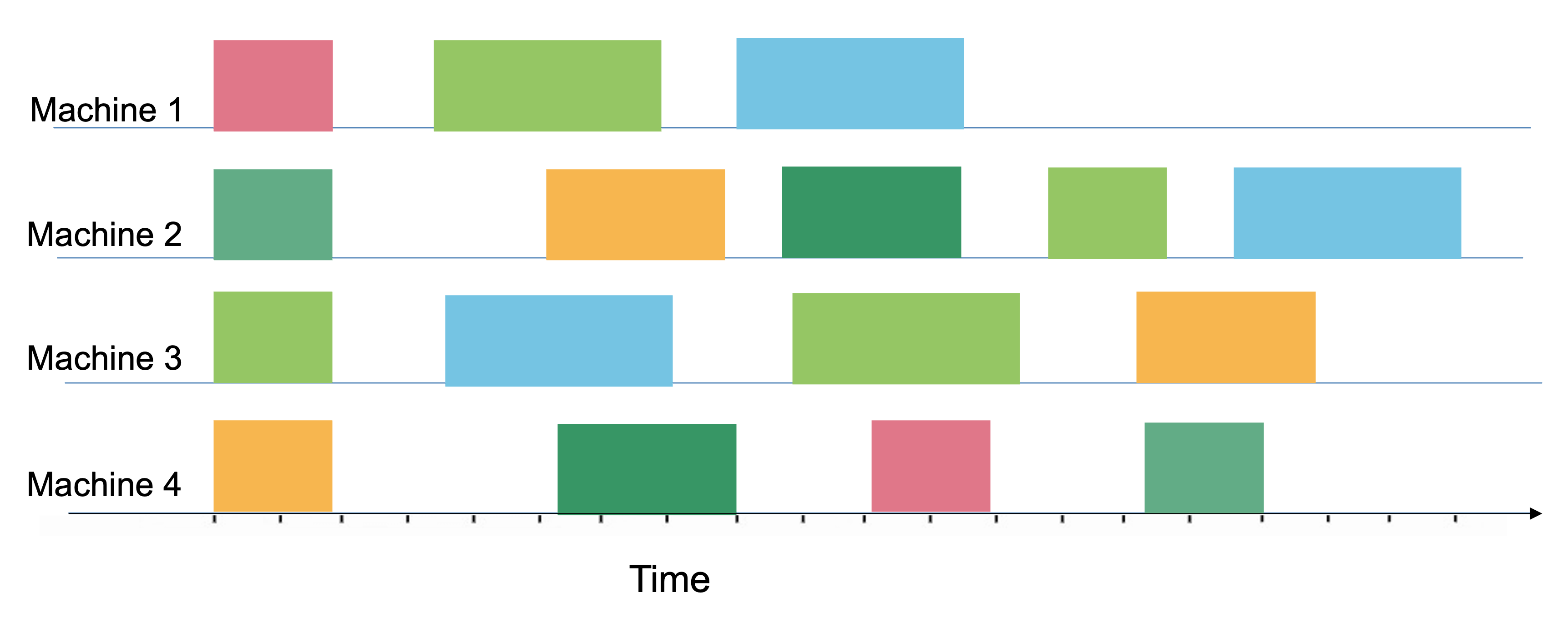}
    \caption{An exemplary solution to a problem with 16 jobs and 4 machines. The idle times of the machines between adjacent jobs depict the required set-up times. The objective is to reduce the total setup time, uniformly distribute the busy times of all machines, and maximize the total value~\cite{PAS}.}
    \label{fig:pas}
\end{figure}

The Production Assignment and Scheduling problem is an NP-hard combinatorial optimization problem involving several jobs and machines and focuses on balancing the processing time on the machines and reducing the manufacturing time of products while maximizing the value of the produced items, as shown in Figure~\ref{fig:pas}.

This optimization task arises, e.g., at business units of BASF. Currently, the optimization process, i.e., finding an optimal or good enough solution, is a time-consuming process and any significant speed-up is highly desirable, especially in the case of an inadvertent breakdown of operations. In the project, quantum annealing and hybrid quantum annealing approaches are tested against the industry scale classical solver.

We develop two distinct mathematical models tailored to the strengths of each solver type. For classical solvers, the problem is formulated as a mixed-integer convex program, leveraging the structure and precision of traditional optimization techniques. In contrast, the quantum-classical approach uses a BQP formulation, which aligns with the capabilities of D-Wave’s Leap Hybrid Solver~\cite{Leap}. This modeling choice ensures that each solver is evaluated using the most appropriate representation of the problem, allowing for a fair and meaningful comparison.

The benchmarking spans problem sizes up to approximately 150,000 variables, a scale that reflects genuine industrial complexity. The study compares an industry-grade classical solver with D-Wave’s hybrid quantum solver, analyzing performance across various metrics. Notably, the quantum-classical solver delivers competitive results, and in some cases, even outperforms the classical solver in terms of speed, demonstrating the potential of hybrid approaches in production environments.

\paragraph{Traffic-Light-based Evaluation}
%For this particular use case, we present all three evaluations, as the results for this use case highlight several aspects of the problem, the model the solver in one or three region of the traffic light spectrum.

\begin{itemize}
\item \textbf{Green Light}
\begin{itemize}
\item \textbf{Scalability}: The results from this use case demonstrate that D-Wave’s hybrid quantum solver can handle problem sizes up to 150,000 variables, which is significant for real-world industrial scheduling. This shows that quantum-classical solvers are not just theoretical but can scale to practical workloads.
\item \textbf{Runtime}: In some benchmark cases, the quantum-classical solver outperformed the classical solver in speed, especially when the problem was modeled in a way that suits quantum optimization (e.g., binary quadratic format). This is a strong indicator of quantum benefit when the problem structure is well aligned.
\end{itemize}
\item \textbf{Yellow Light} 
\begin{itemize}
\item \textbf{Model and Implementation}: The quantum solver’s performance is highly dependent on how the problem is formulated. The binary quadratic model was necessary to solve the problem efficiently with a quantum computer, while the classical solver benefited from a convex formulation. This means quantum benefit is not automatic -- it requires thoughtful modeling.
% \item Solver Selection Trade-offs:
\item \textbf{Solution Quality}: The study shows that while quantum solvers can be competitive, classical solvers still excel in certain structured problems. The benefit of quantum solvers is conditional on problem type and formulation, not universal.
\end{itemize}
\item \textbf{Red Light} 
\begin{itemize}
\item \textbf{Transferability}: The current size of pure quantum hardware does not allow to address limitations like qubit count, connectivity, or noise -- factors that still constrain quantum computing in practice. These could limit applicability in other domains or on different hardware.%\nknote{TODO: Rewrite: transferability to other use cases}
\item No Pure Quantum Benchmark: Since the currently available quantum annealers can not handle the problem sizes which are practically relevant in industry, this work uses D-Wave’s hybrid solver, which combines classical and quantum resources. However, there is no clear information of the amount of quantum computing carried out to get the final results. Hence, it is unclear how much of the performance gain is due to quantum computation alone.
\end{itemize} 
\end{itemize} 

\paragraph{Verdict}
Overall, the work on this use case underscores the importance of problem modeling in unlocking the strengths of both classical and quantum solvers. It provides a compelling case for D-Waves hybrid optimization workflows, especially in domains where traditional methods struggle with scale or non-linearity. 
\begin{itemize}
    \item \textbf{Promising and Practically Viable -- With Caveats}: This use case is well-suited for hybrid quantum-classical approaches. We demonstrated that when the problem is modeled appropriately, e.g., as a binary quadratic program, D-Wave’s hybrid solver can handle large-scale industrial problems (up to ~150,000 variables) and deliver competitive performance, even outperforming classical solvers modeled as a mixed-integer linear program (MILP). This suggests that quantum computing is not just a theoretical possibility but practically viable for certain classes of industrial optimization problems, especially when:
    \begin{itemize}
        \item The problem is non-linear or combinatorial.
        \item Classical solvers face scalability or performance bottlenecks.
        \item The problem can be reformulated into QUBO or BQP, which aligns with quantum solver strengths.
    \end{itemize}
    \item \textbf{However, Quantum Benefit Is Conditional}: The quantum advantage is not universal. It depends heavily on:
    \begin{itemize}
        \item Modeling choices: Classical solvers excel with convex formulations, while quantum solvers need binary/quadratic structures.
        \item Solver maturity: The study uses a hybrid solver, not a pure quantum one, so the quantum contribution is blended with classical resources.
        \item Hardware constraints: The current size of pure quantum hardware does not allow addressing limitations like qubit count or noise, which could affect generalizability, in scaling the problem sizes.
    \end{itemize}
    \item \textbf{Recommendation}: For organizations exploring quantum computing, this use case provides a strong case for hybrid workflows. It is worth investing in:
    \begin{itemize}
        \item Solver-aware problem modeling.
        \item Benchmarking hybrid solvers on real workloads.
        \item Collaborating with quantum vendors to tailor formulations.
    \end{itemize}
\end{itemize}

\subsubsection{Train Routing}

In this use case we investigate the challenge of rescheduling and rerouting trains (hereinafter referred to as \textit{train routing}) in a rail network after disturbances such as equipment failures, environmental hazards, or unexpected delays. Normally, rail schedules are tightly optimized to balance limited resources like tracks and platforms. Once disruptions occur, additional delays emerge as trains compete for these scarce resources. Traditional heuristics and human-driven adjustments often yield workable solutions, but they lack guarantees of optimality. This motivates the exploration of quantum optimization techniques, which can exploit simultaneous state-space exploration to identify globally optimal schedules in exponentially large search spaces.

The problem is formulated first as a MILP, which incorporates real-world constraints such as minimal passing times, single-track exclusivity, station capacities, and headway conditions between trains. Each train’s delay is decomposed into primary (unavoidable) and secondary (avoidable) components, and the optimization objective minimizes the sum of secondary delays weighted by train priorities. Binary precedence variables govern the ordering of trains in shared track segments and binary selection variables determines the selection of track segments to route the trains, while constraints enforce no early departure and minimal station stop requirements. Considering the large scale and high complexity nature of the problem, decomposition techniques have been explored to break the problem into smaller subproblems. In particular, both Lagrangian relaxation and Benders decomposition have been studied as potential approaches to improve computational efficiency. Specifically, with Lagrangian relaxation, each subproblem solves the scheduling and routing of a single train over the network. With Benders decomposition, the delay of each station-train pairs is solved in a master problem, while all the track level coordination, formulated as satisfiability problem, is solved in a subproblem. This significantly reduces the complexity and improves the scalability.

A QUBO model is then derived to map the problem or subproblem after the decomposition to quantum hardware. 
Classical solvers like CPLEX or Gurobi can solve tiny cases in milliseconds and even larger cases below a second. But for large problems as they exist in industry, classical solvers struggle to find the optimum in few seconds, as it is required.

\begin{figure}[htbp]
    \centering
    \includegraphics[width=0.92\linewidth]{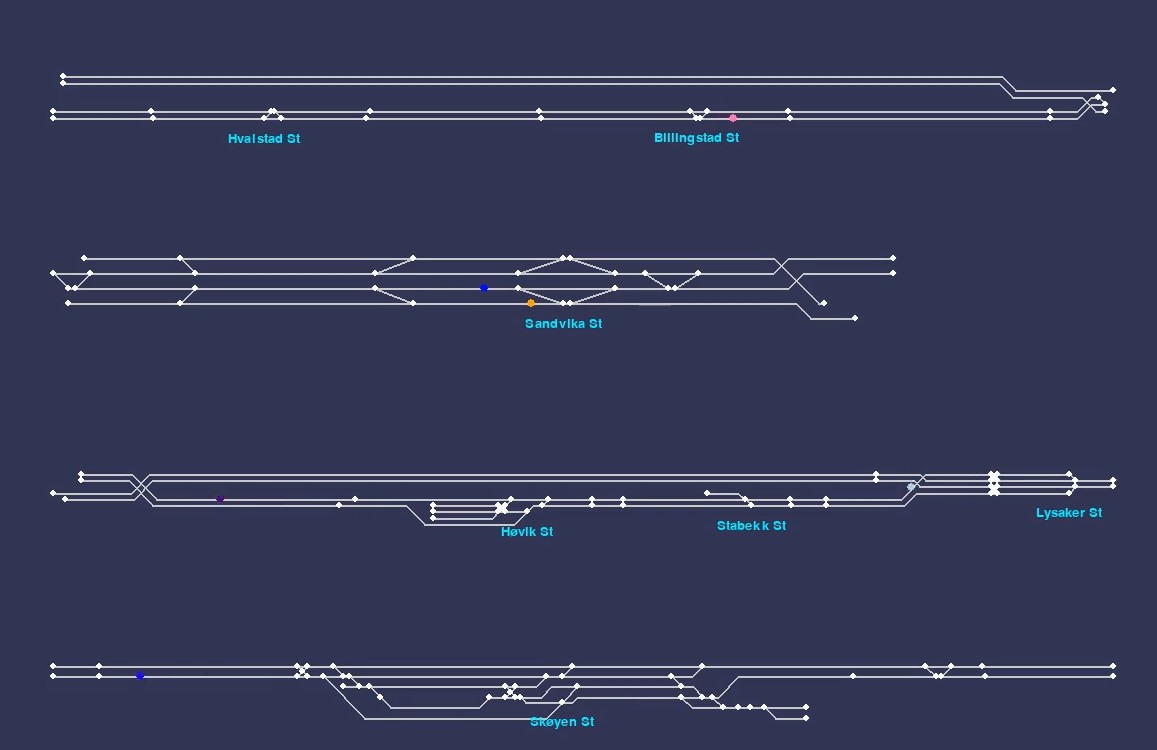}
    \caption{Microscopic illustration of a part of the Norwegian railway network. The colored circles represent the positions of the trains. Visualization based on data from~\cite{railOscope}.}
    \label{fig:routing}
\end{figure}
\paragraph{Traffic-Light-based Evaluation}

\begin{itemize}
    \item \textbf{Yellow Light}
    \begin{itemize}
        \item \textbf{Scalability}:
        To encode the entire problem as a QUBO formulation, the evaluation shows that the number of decision variables scales quadratically with the number of trains and linearly with the number of stations in the worst-case network. This results in cubic scaling when both trains and stations increase, which is acceptable compared to other problems. The scaling matches classical MILP formulations, apart from a constant factor (which is large). However, by applying Benders decomposition, a quantum annealer was able to solve an instance involving 47 tracks, 6 stations and 6 trains using fewer than 240 qubits, while a gate-based method (QAOA+) solved the same instance with fewer than 40 qubits. These results highlight the importance of proper problem decomposition in making quantum approaches more scalable and efficient.
        % The gate-based model achieves solution quality comparable to classical solvers, while the quantum annealer yields solutions of lower quality than those of classical approaches.
        \item \textbf{Transferability}: 
        This use case shares similarities with the sche\-duling and routing of AGV fleets with predefined paths. Therefore, transferability to such use case can be considered in the future research.
    \end{itemize}
    \item \textbf{Red Light}
    \begin{itemize}
        \item \textbf{Runtime}:
        The rescheduling have to be done in seconds in industry. This can be problematic for quantum computing, as there is some overhead due to embedding and pre-/postprocessing steps. Besides, the current results of gate-based quantum hardware takes much longer than the classical computer, while quantum annealing is fast.
        \item \textbf{Solution Quality}:
        Using Benders decomposition, both quantum annealing and gate-based  (QAOA+) quantum computing methods successfully solve a problem instance involving 47 tracks and 6 trains.The gate-based model achieves solution quality close to optimal for those small instances, while the quantum annealer yields solutions
        of lower quality than optimal.
        \item \textbf{Model and Implementation}:
        Current modeling of the use case introduces simplifications such as neglecting the continuous train dynamics, omitting the coordination of transfer options between trains, ignoring the cancellation of trains, and excluding the scheduling of trains at depots. In reality, these factors would introduce additional constraints that increase the complexity of the problem. This further underscores the importance of employing proper decomposition techniques to extract a (sub-)problem structure that is suitable for quantum computing.

    \end{itemize}
\end{itemize}

\paragraph{Verdict}

\begin{itemize}
    \item \textbf{Current Limitation}:
    Attempting to encode the entire problem as a single QUBO formulation turns out to be less efficient and scalable. Currently available hardware is not sufficient to handle real-world instances. Even in the next 10 years, it is unlikely that a system with a sufficiently large number of qubits to fit large real-world instances will be available, given the current roadmaps of the hardware providers. Even when in some point of time hardware is sufficient to embed the problem, it remains to be shown whether the solutions are better than those from classical algorithms. However, it must be emphasised that, due to hardware limitations, only small problem instances can currently be considered. Therefore, based on the investigations carried out, only very limited conclusions can be drawn about future results.
    \item \textbf{Encouraging Findings}: Proper combination of decomposition technique and quantum algorithm is crucial to improve efficiency and scalability. With Benders decomposition, the master problem only encodes the delay of station-train pairs and mathematically equivalent to a set-covering problem, which can be successfully solved by the gate-based quantum algorithm QAOA+ while maintaining a quite low number of Qubits. Its solution quality is comparable to that of classical solvers.
    \item \textbf{Future Potential}:
    The potential future benefit is high, especially since quantum annealing is very fast and well-suited for real-time adjustments, given that the solution quality of quantum annealing will be significantly improved and proper decomposition technique can be developed. 

    \item \textbf{Recommendation}: 
    To improve scalability and efficiency, proper decomposition is essential for solving train routing problems with quantum computing. Combining Benders decomposition and Gate-based methods like QAOA+ shows promising solution quality with minimal Qubit usage, while quantum annealing offers solution speed but currently lower solution quality. Further research work in this direction is recommended.
\end{itemize}

\subsubsection{Pod Coordination}

The pod coordination problem (PCP) models a demand-responsive, door-to-door transport system utilizing passive modular containers (pods) carried by active vehicles such as trains or trucks. This problem falls under the category of active-passive vehicle routing problems with multiple synchronization constraints, necessitating tight coordination between pods and bases. The PCP integrates elements of the dial-a-ride problem, the multi-depot vehicle routing problem, heterogeneous fleets, and multiple modalities, aiming to optimize service quality and operational efficiency. Currently, this use case has not been implemented in a real-life setting, and the complexity of the problem presents significant challenges for both classical and quantum solvers.

\begin{figure}[htbp]
    \centering
    \includegraphics[width=0.92\linewidth]{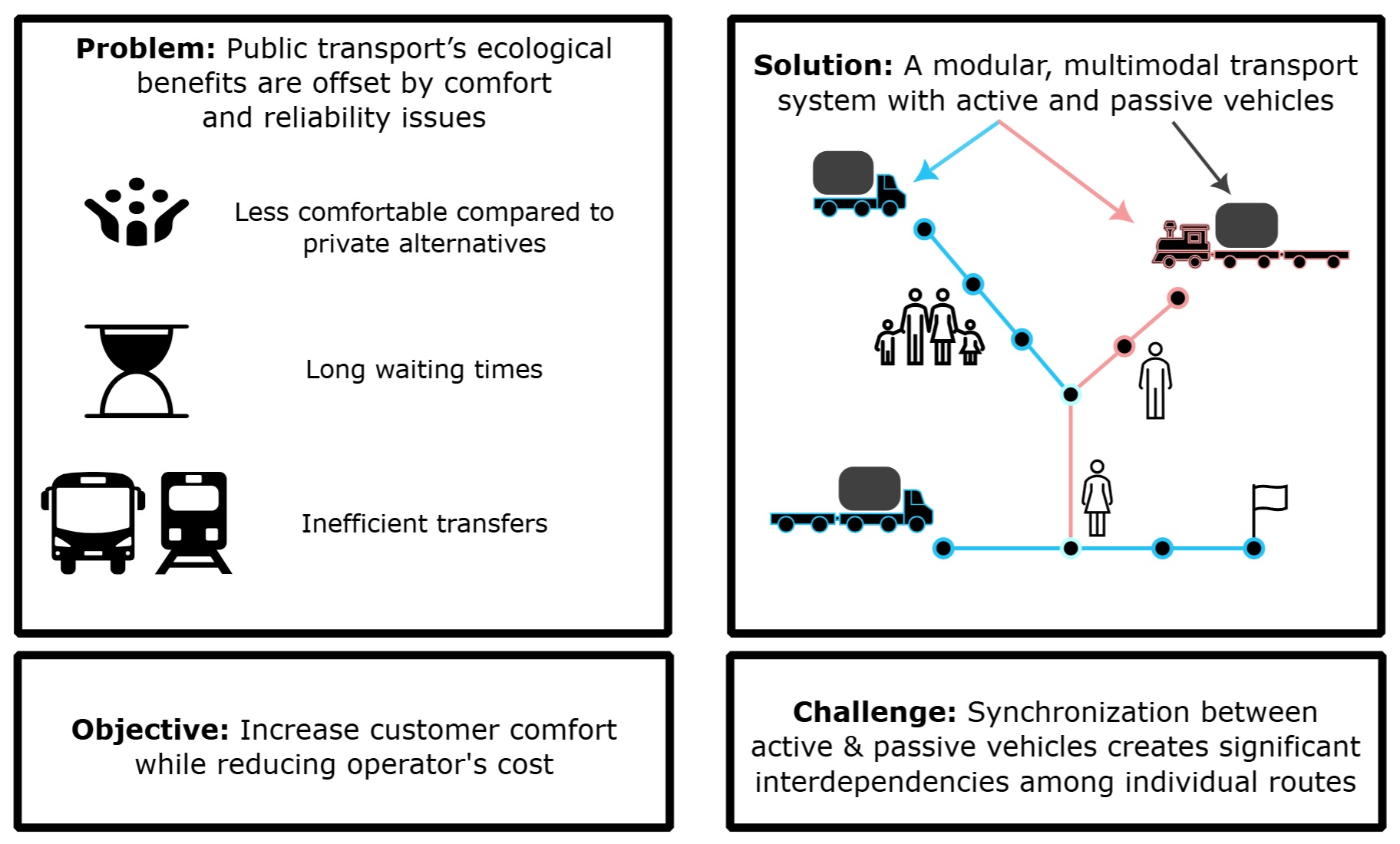}
    \caption{The PCP combines vehicle routing, scheduling, and ordering problems. The goal is to provide an improved experience for users of public transportation services while reducing the operation costs for providers. This figure is adapted from~\cite{Eder2025}. Copyright ©2025 Elsevier.}
    \label{fig:PCP}
\end{figure}

In this project, we explored various classical solution methods, including mixed integer linear programming (MILP), adaptive large neighborhood search, and iterative beam search. The problem's complexity is highlighted by the requirement for a substantial number of qubits; for instance, even the smallest test instances necessitate over 10,000 qubits, making them impractical for current quantum hardware.

\paragraph{Traffic-Light-based Evaluation}

\begin{itemize}
    \item \textbf{Yellow Light}
    \begin{itemize}
        \item \textbf{Model and Implementation}: The theoretical framework for applying quantum computing to the PCP was successfully implemented. However, the need for efficient handling of continuous variables in the MILP formulation, which are required for the synchronization constraints, poses a challenge for quantum computing. Future developments in quantum algorithms that can effectively address these variables will be crucial.
    \end{itemize}
    \item \textbf{Red Light}
    \begin{itemize}
        \item \textbf{Scalability}: The number of qubits required even for the simplest instances (over 10,000 for five customers) far exceeds the capabilities of current quantum hardware. This presents a significant barrier to practical implementation. However, due to these hardware limitations, it is currently impossible to make any reliable statements about scalability with future systems.
        \item \textbf{Solution Quality}: Due to the high qubit requirements, no tests have been performed on quantum hardware. Classical problem-tailored heuristics have demonstrated the ability to find high-quality solutions quickly. This suggests that significant advances in hardware and algorithms are required for quantum approaches to offer a competitive advantage.
        \item \textbf{Runtime}: Due to its intrinsic complexity and high number of hard constraints, embedding the PCP in a quantum circuit currently seems highly inefficient in terms of circuit depth, mapping overhead, and consequently runtime. However, as no tests could be performed on quantum hardware, the current investigations have limited relevance to possible future developments.
        \item \textbf{Transferability}: Even though the PCP represents a rather general problem class and is closely related to the vehicle routing problem and -- in its core -- the traveling salesperson problem, its additional constraints and objectives make it hard to implement on any quantum computer that relies on quantum circuits or a QUBO formulation of the problem.
    \end{itemize}
\end{itemize}

\paragraph{Verdict}

Overall, the pod coordination use case highlights the significant challenges that quantum computing faces in practical applications, particularly in the realm of constraint-heavy optimization related vehicle routing problems.

\begin{itemize}
    \item \textbf{Limited Current Viability}: The high qubit requirements and the inability to conduct experiments on quantum hardware currently render this use case impractical for quantum solutions. The classical methods employed have proven effective, and the existing heuristics outperform quantum approaches at this stage.
    
    \item \textbf{Questionable Future Potential}: Should advancements in quantum hardware and algorithms occur, there may be potential for quantum computing to provide benefits in this area. However, this is highly conditional on overcoming the current limitations of qubit count and the ability to handle hard constraints and continuous variables effectively.
    
    \item \textbf{Recommendation}: Organizations interested in exploring quantum computing for problems related to the pod coordination use case should consider to find models of their problem that rely on discrete variables and fewer (hard) constraints. Otherwise, the PCP is currently not expected to be among the first problem classes that provide a tangible benefit for the industry from advancements in quantum technology within the foreseeable future.
\end{itemize}

% \begin{figure}[htbp]
%     \centering
%     \includegraphics[width=\textwidth]{figures/pods_illustration.pdf}
%     \caption{Initial situation in an example for PCP. Solid (cyan) edges correspond to truck roads, dashed (red) edges to railroads, and the dotted (blue) edge to a ferry route crossing a river. $b_0$ is a truck, $b_1$ is a train, and $b_2$ is a boat. $p_0$ and $p_1$ are pods, and three customers $c_i$ ($i \in \{0,1,2\}$) request transportation from hub $\alpha(c_i)$ to hub $\omega(c_i)$.}
%     \label{fig:pods_2}
% \end{figure}

\subsubsection{Sensor Positioning}
In the automotive industry, freshly produced vehicles need to be transported from the end of the production line to the distribution area. With automated-driving software on board of latest-generation vehicles this task can be done without a human driver. One requirement for this is an environment that is covered by infrastructure-based sensors. The positions of these sensors need to be optimized, such that the covered area is maximized while the number of sensors is minimized.

\begin{figure}[htbp]
    \centering
    \includegraphics[width=0.9\linewidth]{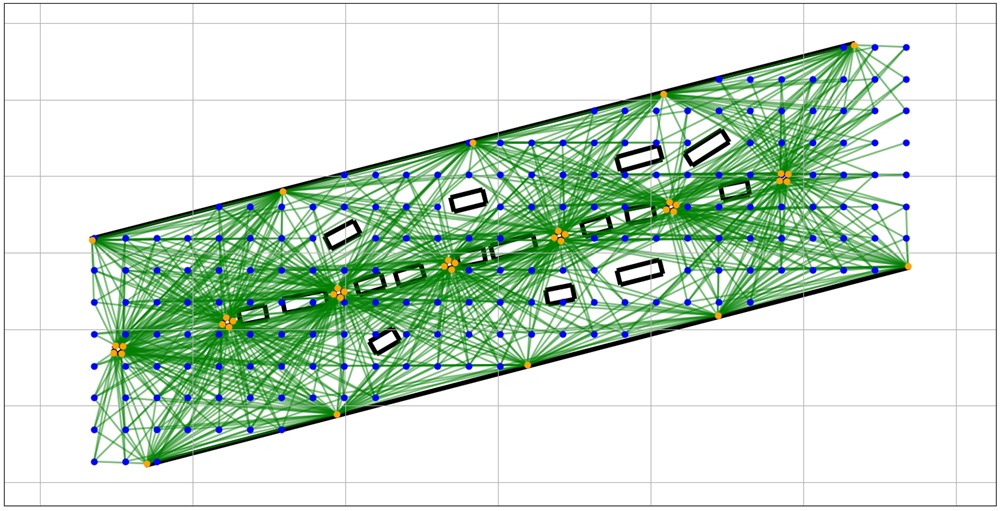}
    \caption{An exemplary scenario with 38 possible sensor positions and 194 street points to cover in a setting inspired by a real-world production environment.}
    \label{fig:sp}
\end{figure}

\begin{figure}[htbp]
    \centering
    \includegraphics[width=0.9\linewidth]{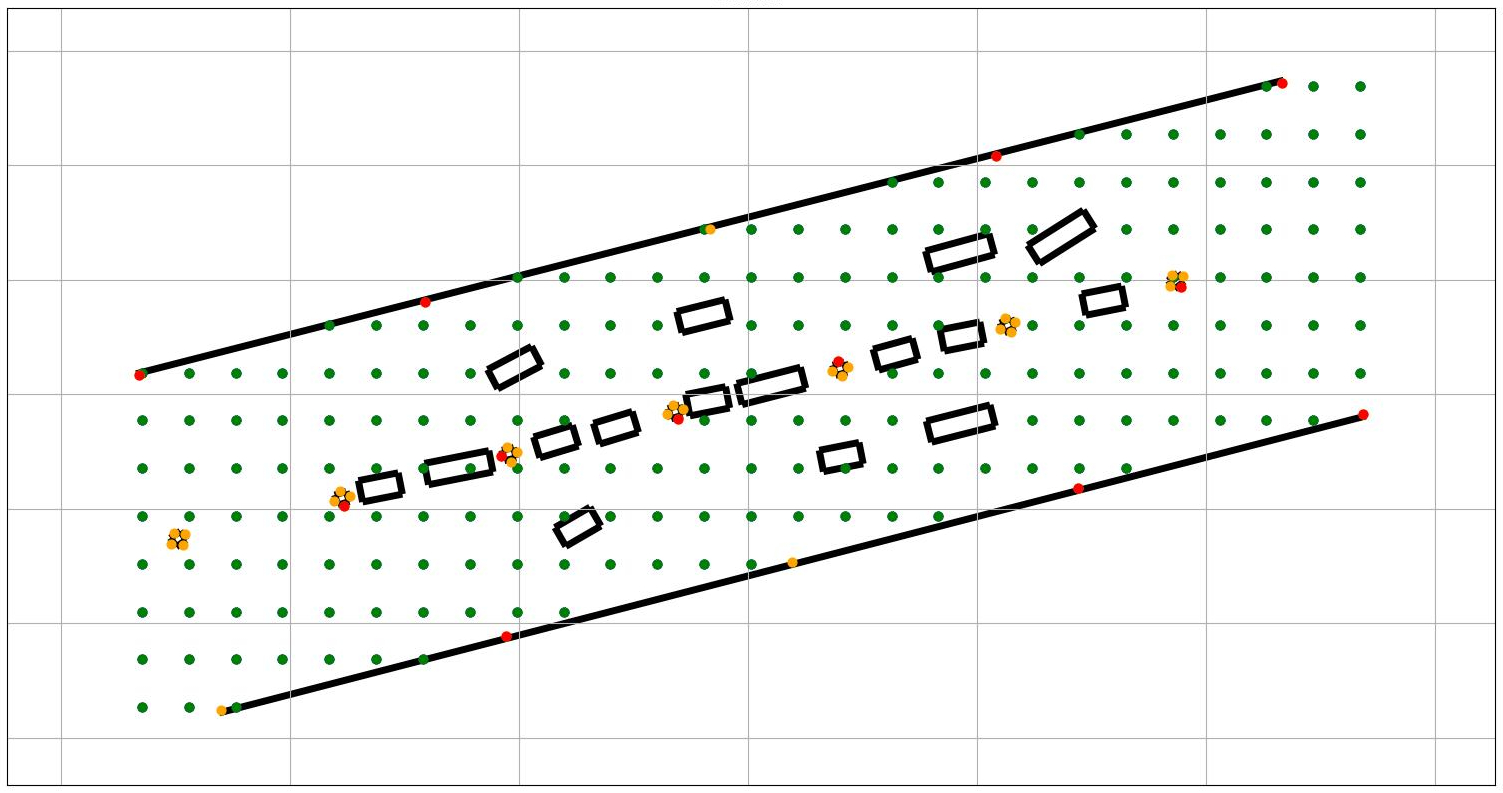}
    \caption{The solution of the example scenario in Figure~\ref{fig:sp} is shown here. 12 sensors are placed on the 38 possible positions and all 194 street points are covered.}
    \label{fig:sp_sol}
\end{figure}

This task can be formulated as a combinatorial optimization problem, typically represented using a QUBO formulation. In this formulation, binary decision variables indicate whether a sensor is placed at a specific location (1) or not (0). The objective function aims to maximize the total coverage area while adhering to constraints related to sensor range and the number of sensors available.

Key components of the formulation include coverage constraints, which ensure that all critical areas are monitored by at least one sensor, and cost considerations that account for installation, maintenance, and operational expenses. Additionally, the model includes structural obstacles like columns that can produce coverage shadows, ensuring that sensors can operate effectively in the intended environment.

%By structuring the sensor positioning problem in this way, researchers can utilize optimization techniques, including quantum annealing, to efficiently explore potential solutions. The ability to model and solve this problem using advanced computational methods highlights the potential of quantum computing in addressing complex real-world challenges.
Hyperparameter optimization of the quantum annealing process includes tuning encoding schemes, penalty factors, chain strengths, and annealing time. Moreover, different decomposition strategies are evaluated, iteratively dividing the model into smaller subproblems, which ensures feasible and scalable solutions for large-scale sensor positioning scenarios. Find more details about this use case in the accompanying paper~\cite{sensor_paper}.

\paragraph{Traffic-Light-based Evaluation}

%For the sensor positioning use case, the evaluation across the traffic light spectrum reflects its current limitations and future potential:

\begin{itemize}
    \item \textbf{Green Light}
    \begin{itemize}
        \item \textbf{Transferability}: The set cover QUBO model we formulated for the sensor positioning use case efficiently identifies minimal coverage solutions, a principle transferable to other domains like test scenario selection for the simulation of driving situations or task-based staffing and scheduling. Its ability to minimize resources while ensuring full coverage makes it adaptable across various automotive and industrial applications requiring optimal coverage with limited assets.
        % \item \textbf{Future Practical Applications}: In the longer term, quantum methods could significantly reduce the number of sensors needed, leading to substantial cost savings and operational efficiency gains.
    \end{itemize}
    \item \textbf{Yellow Light}
    \begin{itemize}
        \item \textbf{Model and Implementation}: The problem fits well into the set cover and maximum independent set frameworks, which are supported by current quantum algorithms and hardware paradigms. Variable counts in classical models are dominated by the sensor positions, while for quantum approaches, we need many slack variables for each constraint.
        \item \textbf{Runtime}: If quantum hardware continues to improve in qubit count, coherence times, and error correction as expected, industry-relevant instances could become solvable within this decade and provide computation times comparable to classical methods according to the test cases we evaluated in this study.
        % \item \textbf{Algorithmic Improvements}: Enhanced embedding techniques and problem-specific quantum algorithms could reduce the required qubits and improve solution stability.
    \end{itemize}
    \item \textbf{Red Light}
    \begin{itemize}
        % \item \textbf{Scalability}: Industry-relevant instances, requiring millions of variables, cannot be tackled with current quantum hardware (e.g., D-Wave Advantage 2 with about 5,000 qubits) without decomposing the problem into subproblems which generally reduces the solution quality.
        \item \textbf{Solution Quality}: Experiments with quantum annealers have produced solutions with low stability and suboptimal quality inferior to classical solvers like Gurobi or heuristics like simulated annealing.
        \item \textbf{Scalability}: For the QUBO formulation of the problem, the number of variables scales nearly cubically with the sensor and street point density, implying that large instances with high-resolution grids of possible sensor position and covered street points will require orders of magnitude more qubits than currently expected available within the next years.
    \end{itemize}
\end{itemize}

\paragraph{Verdict}

The sensor positioning use case illustrates the current limitations of quantum hardware in tackling large-scale, industry-relevant problems. While toy instances with up to 60 qubits have been successfully solved, the solutions are not yet competitive with classical methods in terms of solution quality or stability.

\begin{itemize}
    \item \textbf{Limited Current Utility}: Existing quantum annealing and other quantum approaches are limited to small problem sizes, with solutions often inferior to classical heuristics, and do not justify practical deployment at this stage.
    \item \textbf{Scaling Prospects}: The linear increase in the number of variables with problem size is promising, suggesting that instances with thousands of sensors could become solvable as hardware improves, possibly in the next decade.
    \item \textbf{Future Outlook}: With continued hardware scaling, error correction, and improved embedding techniques, quantum methods could eventually provide cost-effective and optimal sensor placement solutions, reducing infrastructure costs and increasing coverage efficiency.
    \item \textbf{Recommendation}: Organizations should monitor hardware developments and invest in problem modeling strategies that align with quantum capabilities. Collaborations with hardware vendors and research into scalable embeddings are advisable to prepare for future large-scale applications.
\end{itemize}

\subsubsection{Auto-Carrier Loading}
The auto-carrier loading problem (ACLP) is an optimization challenge in the automotive industry, particularly concerning the distribution of new vehicles. While trains are the preferred mode for overland transport, manufacturers often resort to road transport via auto carriers due to flexibility or the unavailability of rail networks. Auto carriers are specialized trucks equipped with flexible loading platforms, and determining the optimal configuration of these platforms is complex due to the varying sizes, shapes, and weights of passenger vehicles.

Typically, a truck and its trailer can carry up to ten vehicles at a time. The vehicles can be arranged on platforms that can be individually angled or combined to maximize space utilization. Additionally, vehicles can be placed either forward or backward to further optimize the load. Constraints related to weight, height, and length must be adhered to for both the truck and trailer, as well as the total load.

In theory, the solution space for the ACLP scales exponentially with the number of vehicles, leading to a vast number of possible arrangements~\cite{Jaeck_ACL}. However, the objective is highly discretized, meaning that while many arrangements may meet the requirements, identifying an optimal solution can be relatively straightforward due to the potential for simple heuristics or human decision-making to yield satisfactory results.

In this project, we explore various classical solution methods, including simulated annealing and CPLEX~\cite{cplex}, a commercial optimization software, alongside the D-Wave quantum annealer~\cite{d-wave-advantage}. The complexity of the problem is highlighted by the requirement for a substantial number of variables; for instance, toy problems have been investigated with a pool of up to 18 cars placed on up to six platforms, leading to QUBO matrices with up to 1000 variables. However, the currently available quantum hardware is insufficient to solve industry-scale problem instances.

\begin{figure}[htbp]
    \centering
    \includegraphics[width=0.82\linewidth]{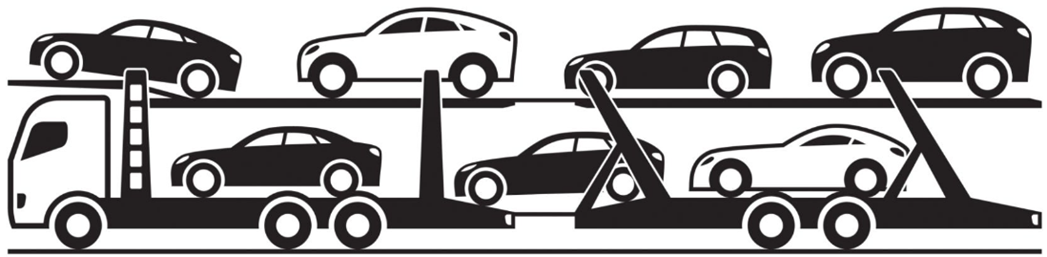}
    \caption{Example solution for loading seven vehicles on a truck and a trailer.}
    %Example solution for loading seven vehicles on a truck and trailer.
    \label{fig:acl}
\end{figure}

\paragraph{Traffic-Light-based Evaluation}

%For the auto-carrier loading use case, we present evaluations across the traffic light spectrum, reflecting the challenges and limitations faced in applying quantum computing to this problem.

\begin{itemize}
    \item \textbf{Green Light}
    \begin{itemize}
        \item \textbf{Model and Implementation} The theoretical framework for applying quantum computing to the ACLP exists. If hardware limitations are overcome, quantum methods could potentially optimize complex loading problems more effectively than classical methods.
    \end{itemize}
    \item \textbf{Yellow Light}
    \begin{itemize}
        \item \textbf{Scalability \& Runtime}: The problem scales linearly with the number of cars. This holds for classical MILP formulations as well as for QUBO formulations, which makes quantum computing promising. However, the problem has a lot of constraints, some of them of fourth order. This leads to many slack variables in the QUBO formulation, which means that the number of variables for quantum computing is higher by a constant but large factor. The currently available quantum hardware is not sufficient to solve industry-scale problem instances and it remains to be seen if quantum methods can compete with runtimes achieved with classical heuristics that produce sufficient solution qualities. % \menote{TODO: One sentence about the linear scaling of variables} The large polynomial count of the optimization problem complicates efficient encoding into a quadratic equation, which is necessary for quantum approaches. \nknote{Do you mean the constraints with fourth power equations? We even need to make them linear to then put it into quadratic penalty terms.}
        \item \textbf{Solution Quality}: The D-Wave quantum annealer has provided results for small problem instances comparable to solutions found with classical heuristics like simulated annealing, indicating that the solution quality could also be competitive for larger problem sizes once quantum hardware matures.
        \item \textbf{Transferability}: If advancements in quantum hardware occur, there may be potential for quantum computing to optimize entire fleets of trucks at a level of detail currently only possible for individual auto carriers.
    \end{itemize}
\end{itemize}

\paragraph{Verdict}

The question if there is a clear path to quantum advantage in the ACLP within the near future is delicate. The implications would be significant for related problems in logistics in the automotive sector and beyond. 

\begin{itemize}  
    \item \textbf{Encouraging Findings}: Empirical observations suggest that the number of model variables increases linearly with the number of cars included in the problem. This linear relationship implies that problems involving up to 100 cars and 10 trucks could potentially be addressed with approximately 100,000 noisy qubits, meaning real-world-related problem sizes could be solved with quantum devices within a matter of years.

    \item \textbf{Limited Potential for Improvement}: While the D-Wave quantum annealer has been able to solve small problem instances, the results obtained were not significantly better than those achieved through classical heuristic solutions. This indicates that it might be hard for quantum computing to provide a competitive advantage for this specific application due to the relatively good performance of classical methods.
    
    \item \textbf{Recommendation}: Until quantum hardware is mature and provides the necessary qubit counts to solve large-scale problems, industries that rely on efficient logistics and face challenges similar to the ACLP should explore the integration of quantum solutions into existing logistics systems. This would accelerate the transition to improved decision-making processes that have a huge scaling potential due to the high number of vehicles and trucks involved in the logistic frameworks of globally acting manufacturers.
\end{itemize}

% ###########################################################

\subsubsection{Modular Production Logistics}
\begin{figure}[htbp]
    \centering
    \includegraphics[width=0.8\textwidth]{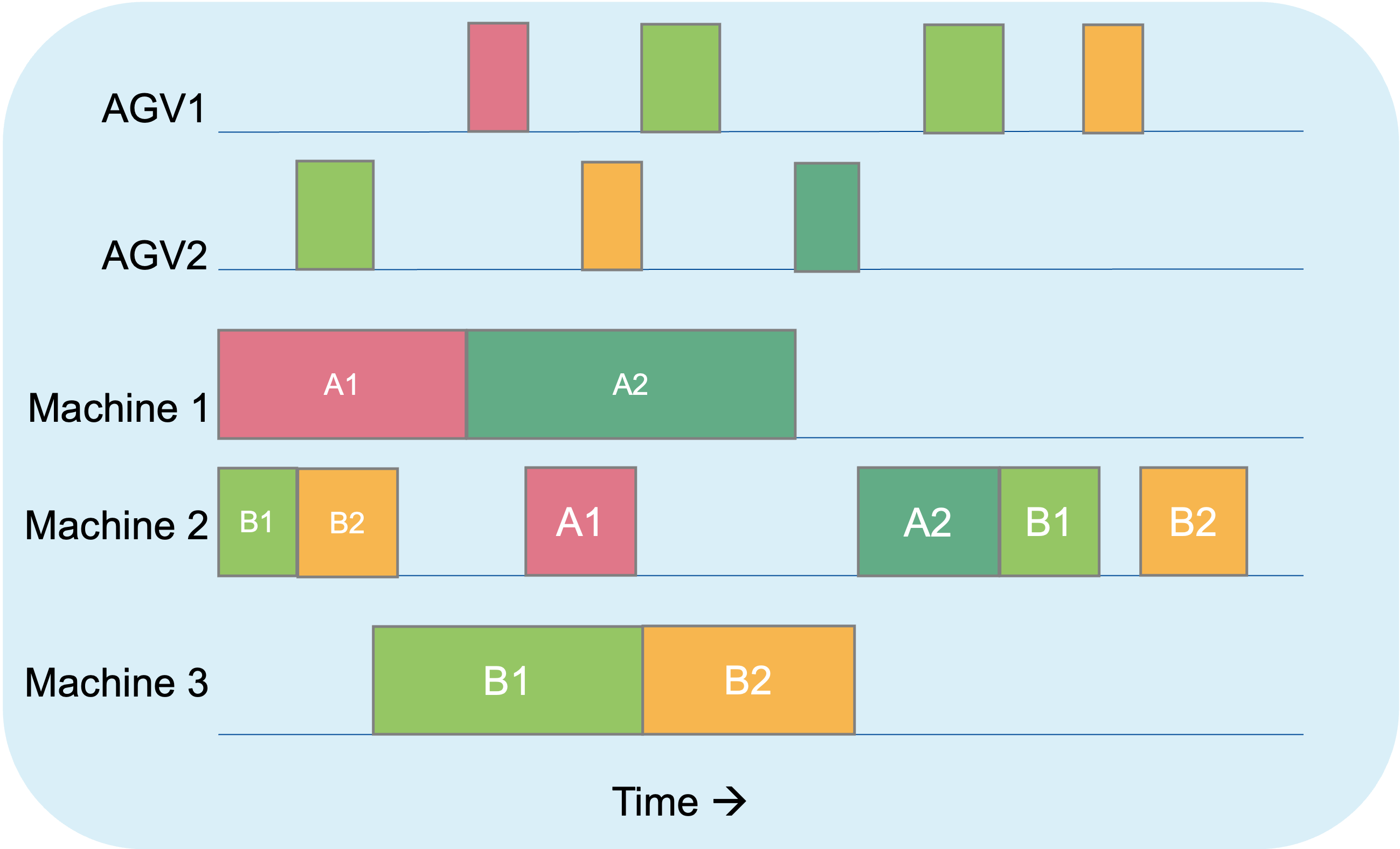}
    \caption{A schematic diagram of problem instance with two AGVs, total 2 jobs and 3 machines. The transportation of any job from one machine to another is carried out by an AGV.}
    \label{fig:moprolog}
\end{figure}

In this use case we study a complex industrial scheduling problem involving autonomous guided vehicles (AGVs) in a modular logistics setting, and present a comparative study of quantum-classical and classical optimization methods. Figure~\ref{fig:moprolog} shows an illustration of a small process involving different jobs and AGVs.

The use case arises at several filling floors at BASF, where pallets are filled with different products and are then wrapped for shipping. In this project, we evaluate how different modeling approaches affect solver performance and to explore whether quantum-classical metaheuristics can offer practical advantages over traditional methods.

Two distinct mathematical models are developed for the same scheduling problem. The first is a time-indexed MILP, which is well-suited for classical mathematical programming solvers. The second is a binary optimization model with both linear and quadratic constraints and a linear objective function, designed to be compatible with quantum-classical metaheuristics. This dual-modeling approach allows us to assess solver performance not just across technologies, but also across formulations tailored to each solver’s strengths.

Our study finds that modeling choices significantly influence optimization outcomes. Classical solvers perform well with the MILP formulation, while the quantum-classical metaheuristic shows strong results with the binary quadratic model. This highlights the importance of aligning problem structure with solver capabilities. The quantum-classical approach, in particular, benefits from the novel modeling technique, suggesting that quantum-inspired methods can be competitive in real-world industrial contexts when the problem is appropriately framed.

Overall, the results obtained using classical computing are far better than quantum computing. However, this use case provides a compelling evidence that quantum-classical metaheuristics are a viable tool for industrial scheduling, especially when traditional solvers face limitations mainly due to ill-modeled mathematical formulations of the optimization problem. It thus reinforces the broader insight that solver-aware modeling is essential to unlock the full potential of any computational paradigm, classical or quantum. 

\paragraph{Traffic-Light-based Evaluation}
%This use case definitely does not qualify for a green-light evaluation as ultimately with better modeling on classical computing one can obtain much better results than quantum computing. However, this use case does offer some encouraging insights. 

\begin{itemize}
\item \textbf{Yellow Light}

\begin{itemize}
\item \textbf{Model and Implementation:} The quantum-classical meta\-heu\-ristic solver performs well when the problem is modeled as a binary optimization problem with quadratic constraints. This formulation aligns with the strengths of hybrid classical-quantum solvers, allowing them to deliver competitive results. However, this benefit is not inherent to quantum computing—it is highly dependent on the modeling strategy. The classical solver, using a time-indexed MILP formulation, outperform the hybrid solvers. This shows that quantum benefit is conditional, and solver performance is tightly coupled to how the problem is framed.
\item \textbf{Scalability:} The quantum-classical solver used is a metaheuristic, which means it does not guarantee optimality. While it shows promising results, its performance is instance-specific and may vary depending on the complexity and constraints of the scheduling task. This limits its predictability and reliability in broader applications without further tuning.
\item \textbf{Solution Quality:} The D‑Wave hybrid solver produced feasible solutions for the scheduling problem, but the solution quality was noticeably lower than that achieved by state-of-the-art classical solvers like Gurobi. However, the results are far from negligible—they demonstrate that hybrid quantum solvers can tackle complex optimization tasks and provide a foundation for future improvements.
\end{itemize}

\item \textbf{Red Light} 

\begin{itemize}
\item \textbf{Transferability:} While the results are promising for the specific AGV scheduling use case, there is limited evidence of generalisability. It remains uncertain whether a quantum-classical benefits would be observed in other logistics optimization problem without comparable modeling adaptations.
\item \textbf{Runtime}: The hybrid solver achieved feasible solutions for the scheduling problem but generally required longer total runtimes than Gurobi. While Gurobi consistently solved instances to near-optimality within seconds or minutes, the hybrid approach showed higher variability in runtime, sometimes taking several minutes for complex cases.
\item No Pure Quantum Solution: Due to proprietary nature of D-Waves hybrid solvers, we could not assess the actual effect of quantum computing alone, so it’s unclear how much of the performance gain is attributable to quantum computation versus classical heuristics embedded in the hybrid solver. This makes it difficult to assess or quantify for a true quantum advantage. Additionally, the size of the hardware restricts us to address hardware constraints such as qubit count, connectivity, or noise—factors that are critical in assessing the scalability and feasibility of quantum solutions in real-world settings.
\end{itemize}

\end{itemize}

\paragraph{Verdict}

\begin{itemize}
\item \textbf{Quantum-Classical Optimization Is Promising}: Binary formulations with quadratic constraints allow hybrid solvers to perform competitively, showing that quantum-inspired methods can be applied to real-world industrial challenges, especially when problems are modeled to suit quantum paradigms.

\textbf{Quantum Advantage Is Conditional and Model-dependent}: Classical solvers using time-indexed MILP outperform hybrids when the problem structure leads to linear or convex formulations. Benchmarking against pure quantum solvers was not feasible due to hardware limitations.

\item \textbf{Quantum-Classical Approaches Are Complementary to Classical Methods}: While effective in certain scenarios, they are not replacements. Classical solvers remain robust and reliable, particularly when the problem structure aligns with their strengths, and quantum benefits are not yet broadly validated.
\end{itemize}

In summary, the work on this use case shows that quantum-classical metaheuristics can be effective for industrial scheduling -- but the quantum benefit is conditional, modeling-dependent, and not yet broadly validated. Classical solvers remain robust and reliable, especially when the problem structure suits their strengths.
\begin{itemize}
\item \textbf{Recommendation}: For organizations exploring quantum computing, this use case provides a strong case for hybrid workflows. It’s worth investing in:
\begin{itemize}
\item Invest in solver-aware modeling frameworks:
The study shows that quantum-classical solvers perform well when the problem is modeled as a binary optimization problem with quadratic constraints. Future efforts should focus on developing automated or semi-automated modeling pipelines that can translate industrial scheduling problems into quantum-suitable formats like QUBO or BQP. This will reduce the manual effort and expertise required to unlock quantum benefits.
\item Benchmark against pure quantum solvers:
To better understand the true quantum contribution, future work should include comparative benchmarks with pure quantum solvers (e.g., gate-based systems or annealers without classical augmentation). This will help isolate quantum performance and guide decisions on when hybrid vs. pure quantum approaches are most appropriate.
\item Explore generalizability cross logistics domains:
The current study focuses on AGV scheduling in modular logistics. Future research should test quantum-classical metaheuristics on other logistics problems—such as warehouse routing, supply chain optimization, or multi-modal transport planning—to assess how broadly the modeling and solver strategies apply.
\item Integrate hardware-aware constraints:
Quantum hardware limitations (e.g., qubit count, connectivity, noise) are not addressed as the hardware size is extremely restrictive for this use case. Future implementations should incorporate hardware-aware modeling and scheduling, ensuring that problem instances are tailored to the capabilities of available quantum devices. This will improve feasibility and performance in real deployments.
\end{itemize}
\end{itemize}

\subsubsection{Cooling System Optimization}

The Cooling System Optimization problem addresses the challenge of designing an optimal architecture for mechatronic systems composed of different components (e.g., batteries, engines, coolers, pumps). The task involves deciding which components should be included and how they should be interconnected under given operational conditions and constraints (e.g., weight, cost, or thermal limits). Evaluating system performance requires simulating the physical behavior of the components, which is modeled by (differential) equations. The problem combines elements of optimization and simulation and falls into the academic area of Optimization of Dynamic Systems (closely related to Optimal Control). The industrial goal is to balance system performance with cost and efficiency in a way that is computationally tractable. The system is defined by a set of differential equations which can be linearized into a linear system of equations. The connections between the components are represented by binary variables which signal if a component is connected or not. These can be integrated into the linear system by multiplying the relevant equation with the corresponding decision variable (i.e. connection). Optimal solutions then are represented by the minimal assignment of the connection variables that optimizes a certain metric (e.g., the temperature in one or multiple components).

\begin{figure}[ht]
    \centering
    \includegraphics[width=0.8\linewidth]{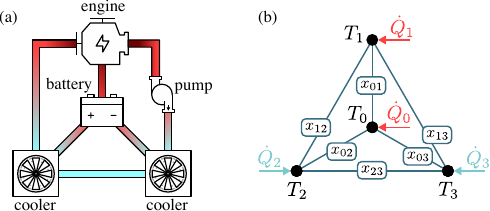}
    \caption{Simple example of a cooling system and its mathematical abstraction into a linear system.}
    \label{fig:cooling}
\end{figure}

The test instances scale from minimal setups (four components and six possible connections) to industrially relevant cases with around ten components under fifty operating conditions. Classical methods currently rely on genetic algorithms combined with explicit integration schemes, with runtimes extending to several hours for larger cases. For the quantum approach, the use case explores the Quantum Approximate Optimization Algorithm (QAOA) for optimization, alongside Quantum Singular Value Transformation (QSVT) for simulation. Since the circuit depth of QSVT is prohibitive for current NISQ devices the approach was simulated only and not tested on actual quantum hardware. More information can be found in the paper detailing the implementation and evaluation \cite{hölscher2025cooling}.

\paragraph{Traffic-Light-based Evaluation}
\begin{itemize}
    \item  \textbf{Green Light}
    \begin{itemize}
        \item \textbf{Transferability}: The approach could be extended to other use cases where linear systems form the core simulation step (e.g., unit commitment or topology optimization).
    \end{itemize}
    \item \textbf{Yellow Light}
    \begin{itemize}
        \item \textbf{Solution Quality}: Given sufficient resources the algorithm was able to find the optimal solution for very small systems. Larger systems could not be simulated due to the complexity of the algorithm so the effectiveness of QAOA remains to be tested. Using Grover would render an algorithm with provable effectiveness.
    \end{itemize}
    \item \textbf{Red Light}
    \begin{itemize}
        \item \textbf{Scalability}: For the optimization each possible connection is represented by a qubit. In the worst case (all-to-all connectivity) he number of connections scales quadratically with the number of components. Current NISQ devices cannot support QSVT due to noise and circuit depth; logical qubits on fault-tolerant machines are required. 
        \item \textbf{Runtime}: The actual extent of runtime advantage achievable depends on an efficient encoding of the system (i.e. matrix and solution vector) and is not demonstrated for general realistic scenarios.
        \item \textbf{Model and Implementation}: Modeling the use case as a linear system limits the complexity of the dynamical system to systems without non-linear elements which is not sufficient for industrial problems. 
    \end{itemize}
\end{itemize}

\paragraph{Verdict}

The cooling system optimization case is assessed as Red in the traffic-light evaluation. 
\begin{itemize}
    \item \textbf{Open Challenges}: Although the theoretical potential for runtime speedups is significant—particularly in solving linear systems—the approach is currently limited by both hardware readiness and algorithmic barriers. The restriction to linear systems limits the usefulness of the approach for the specific use case of optimizing dynamical systems. In general an efficient encoding of the system matrix is not given and reduces any potential speedup. This represents a challenge independent of the current hardware capabilities. As the approach utilizes fault-tolerant algorithms hardware limitations however still negate current applicability.
    \item \textbf{Transferable Findings}: The applicability to problems defined by linear systems holds great value in many other use cases, especially in solving numerical simulations of physical quantities which have an efficient classical approach (e.g., Finite-Element Method, Finite Volume Method, etc.) to discretizing the problem into linear systems. Further utilizing the result from a quantum linear system solver within an optimization algorithm like QAOA or Grover, can remove the bottleneck of reading out the full state vector and making use of solving the linear system in superposition. 
    \item \textbf{Recommendation}: Continued exploration is justified, especially in contexts where linear formulations are natural, but breakthroughs in both algorithm design and hardware capabilities are prerequisites for industrial deployment. Especially an efficient way to improve the scaling of the condition number (influencing the complexity of QSVT) and an efficient encoding are important challenges to overcome.
\end{itemize} 

\subsection{Machine Learning Use Cases}
\label{sec:QML_use_cases}
\subsubsection{Cart Pole Quantum Twin}

\begin{figure}
    \centering
    \includegraphics[width=0.5\linewidth]{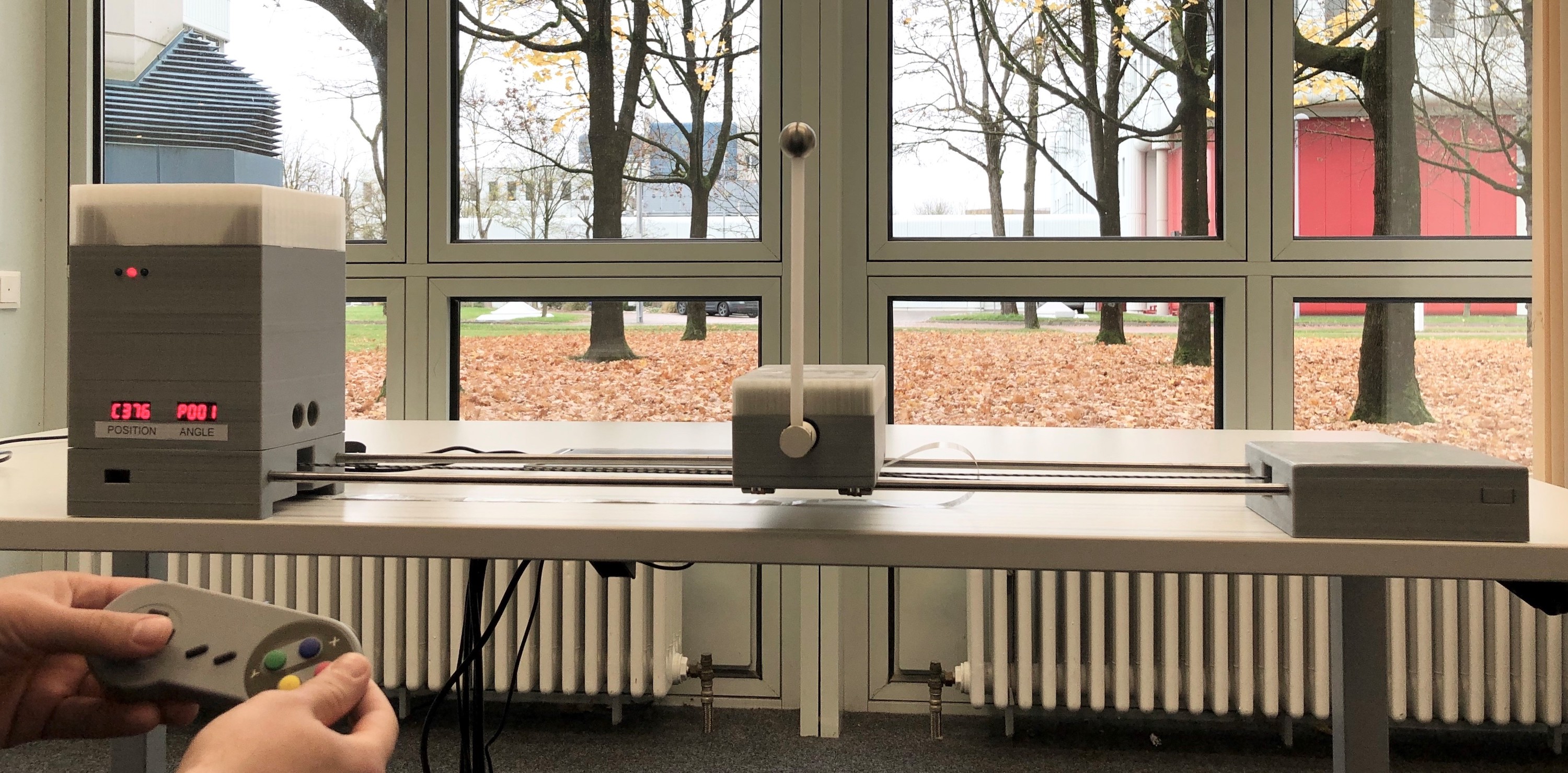}
    \caption{Photo of a physical realization of a cart pole.}
    \label{fig:cart-pole}
\end{figure}

The Cart Pole Quantum Twin problem is based on the well-known cart-pole or inverted pendulum benchmark from control engineering. While the system is very low-dimensional (four variables), it is nonlinear and non-trivial, making it a representative case for general control tasks. In this project, the cart-pole serves as a proxy for more complex industrial control problems (e.g., gas turbine control with hundreds or thousands of sensors), while remaining small enough to be simulated on today’s quantum simulators. The use case was investigated with quantum machine learning approaches and explores whether quantum-enhanced training methods can provide benefits over classical approaches in designing control policies.

The toy instance can be represented using four qubits, while realistic industrial benchmarks (e.g., the industrial benchmark by Hein et al. 2017) would likely require 30–50 qubits. The key performance criterion is solution quality, with secondary attention to computation time. In practice, quantum solutions achieved nearly comparable quality to classical methods, but were 10–100 times slower. Classical baselines included neural networks optimized with particle swarm optimization. On the quantum side, experiments used variational quantum circuits (VQCs), quantum long short-term memory (LSTM), and quantum–particle swarm hybrids. All implementations were conducted on simulators via Qiskit, with no results on actual quantum hardware. 

The proof-of-concept confirmed that the approach works on toy problems but did not extend beyond them. The main advantage anticipated for quantum approaches lies in their ability to explore the solution space without relying on gradients and without being trapped in local minima. Current NISQ hardware is insufficient due to noise and depth limitations, but the path forward involves scaling to noiseless 30–50 qubit systems with circuit depths of around 20 VQC layers for simulation and thousands for optimization. The expectation is that future fault-tolerant devices will be required for meaningful industrial-scale experiments. No major unforeseen blockers were encountered during implementation. For further details the use case and experiments are published in \cite{sun2025}.

\paragraph{Traffic-Light-based Evaluation}

\begin{itemize}
    \item  \textbf{Green Light}
    \begin{itemize}
        \item \textbf{Solution Quality}: Simulations show comparable results compared to classical machine learning methods — better quantum hardware should enable larger and more realistic benchmarks.
        \item \textbf{Transferability}: Transfer opportunities exist for real-world control problems such as plasma stabilization in fusion reactors, wind turbine regulation, design optimization, and even drug discovery.
    \end{itemize}
    \item \textbf{Yellow Light}
    \begin{itemize}
        \item \textbf{Scalability}: It is not yet known how well quantum methods will scale to more realistic benchmarks beyond toy models, nor what requirements are posed on coherence time and noise to facilitate training on actual quantum hardware.
        \item \textbf{Model and Implementation}: Quantum machine learning models are similarly applicable compared to classical machine learning models and featured comparable performance for small instances. Scalability to larger instances still needs to be proven on larger hardware.
    \end{itemize}
    \item \textbf{Red Light}
    \begin{itemize}
        \item \textbf{Runtime}: Current quantum simulations are significantly slower (10–100x) than classical methods. Without speedups, practical relevance remains limited.
    \end{itemize}
\end{itemize}

\paragraph{Verdict}

The Cart Pole Quantum Twin is assessed as Green in the traffic-light evaluation. The path to demonstrating potential quantum advantage appears clearer than for many other use cases, as the problem is well defined, scalable in principle, and closely linked to industrial relevance. 
\begin{itemize}
    \item \textbf{Reliance on Future Hardware}: Meaningful results depend entirely on future advances in fault-tolerant quantum hardware due to requirements on circuits with more layers than current hardware can handle without accumulating to many errors. 
    \item \textbf{Optimistic Outlook and Applicability}: The main appeal lies in the possibility of optimization of quantum circuits to avoid local minima and explore larger solution spaces, potentially yielding better-than-classical controllers. Transfer potential is broad, suggesting this line of research could eventually support high-impact industrial applications once sufficient hardware capabilities are available.
    \item \textbf{Recommendation}: Closely monitor hardware developments and test problem instances with higher complexity on improved quantum processors.
\end{itemize}

\subsubsection{Contextual Bandit}

The Contextual Bandit problem involves identifying an optimal decision in a high-dimensional space where the best action depends on external contextual information, which itself may also be high-dimensional. This problem often arises in optimal control of industrial machines and systems. Deciding on the best control input depends on many external variables (in complex cases hundreds or more) and thus finding the best option that maximizes a certain goal is non-trivial. In machine learning, this is referred to as the continuous action contextual bandit problem. For this project, the Industrial Benchmark \cite{Hein_2017} was used as a representative environment, designed to capture realistic aspects of industrial control scenarios. In this framing, optimization requires searching over control variables while keeping them fixed, and then measuring system performance in response. This use case falls under machine learning and aims to assess whether quantum approaches can improve search efficiency in large, complex spaces.

The initial experiments were conducted on toy instances modeled with five qubits. Classical baselines used neural networks, while quantum variants were implemented using variational quantum circuits (VQCs) with TensorFlow Quantum. While practical limitations concerning computational resources and existing hardware persist, the robust generalization performance of VQCs, combined with rapid progress in quantum technologies, highlights their potential for industrial optimization tasks, particularly in high-dimensional settings and underscores the need for further exploration within quantum machine learning. For industry-scale problems, the number of required qubits and performance characteristics remain unknown.

The experiments demonstrated that the contextual bandit formulation can be mapped to a small quantum system but did not extend beyond the toy case. No runs were performed on real hardware; instead, classical workstations were used to simulate quantum circuits. While the potential of quantum contextual bandit solvers is high—especially in domains where exhaustive exploration of vast solution spaces is required—the practical challenges remain unresolved. These include how to handle large data volumes, what is the best ansatz for a specific problem, and how to integrate a quantum circuit based bandit into real-world workflows. More details on the benchmark problem and performed investigations can be taken from~\cite{Schulte2025}

\paragraph{Traffic-Light-based Evaluation}

\begin{itemize}
    \item  \textbf{Green Light}
    \begin{itemize}
        \item \textbf{Transferability}: Contextual bandits underpin numerous applications, from recommendation systems and online advertising to clinical decision support, robotics, and financial optimization.
    \end{itemize}
    \item \textbf{Yellow Light}
    \begin{itemize}
        \item \textbf{Solution Quality}: The quality of the solutions from the quantum circuits show competitive performance and even feature a better generalization compared to classical models.
    \end{itemize}
    \item \textbf{Red Light}
    \begin{itemize}
        \item \textbf{Runtime}: Industrial instances of the contextual bandit often have real-time constraints which pose challenges on the integration of quantum hardware into workflows. Current quantum simulations are significantly slower (10–100x) than classical methods. Without speedups, practical relevance remains limited. 
        \item \textbf{Scalability}: The quantum resource requirements for real-world contextual bandit problems are not yet quantified, leaving uncertainty about whether practical quantum advantage is attainable.
        \item \textbf{Model and Implementation}: Efficient encodings of real-world data are not known for general cases.
    \end{itemize}
\end{itemize}

\paragraph{Verdict}
The Contextual Bandit use case is assessed as Amber in the overall traffic-light evaluation. 

\begin{itemize}
    \item \textbf{Reliance on Future Hardware}: On current hardware no usable solutions could be generated and thus future hardware has to improve to yield comparable results to the performed simulations. 
    \item \textbf{Applicability}: Theoretical impact is high — given the ubiquity of contextual bandits across industries — but practical application requires both better hardware and new modeling strategies. 
    \item \textbf{Recommendation}: Expanding tests beyond toy problems by increasing context dimensionality and means for efficiently encoding real-world data.
\end{itemize}

\subsubsection{Forecasting}

Forecasting timeseries is a very important topic in many areas, such as finance, logistics (e.g., commodity price forecasting) or other areas like weather prediction.

\begin{figure}
    \centering
    \includegraphics[width=0.8\linewidth]{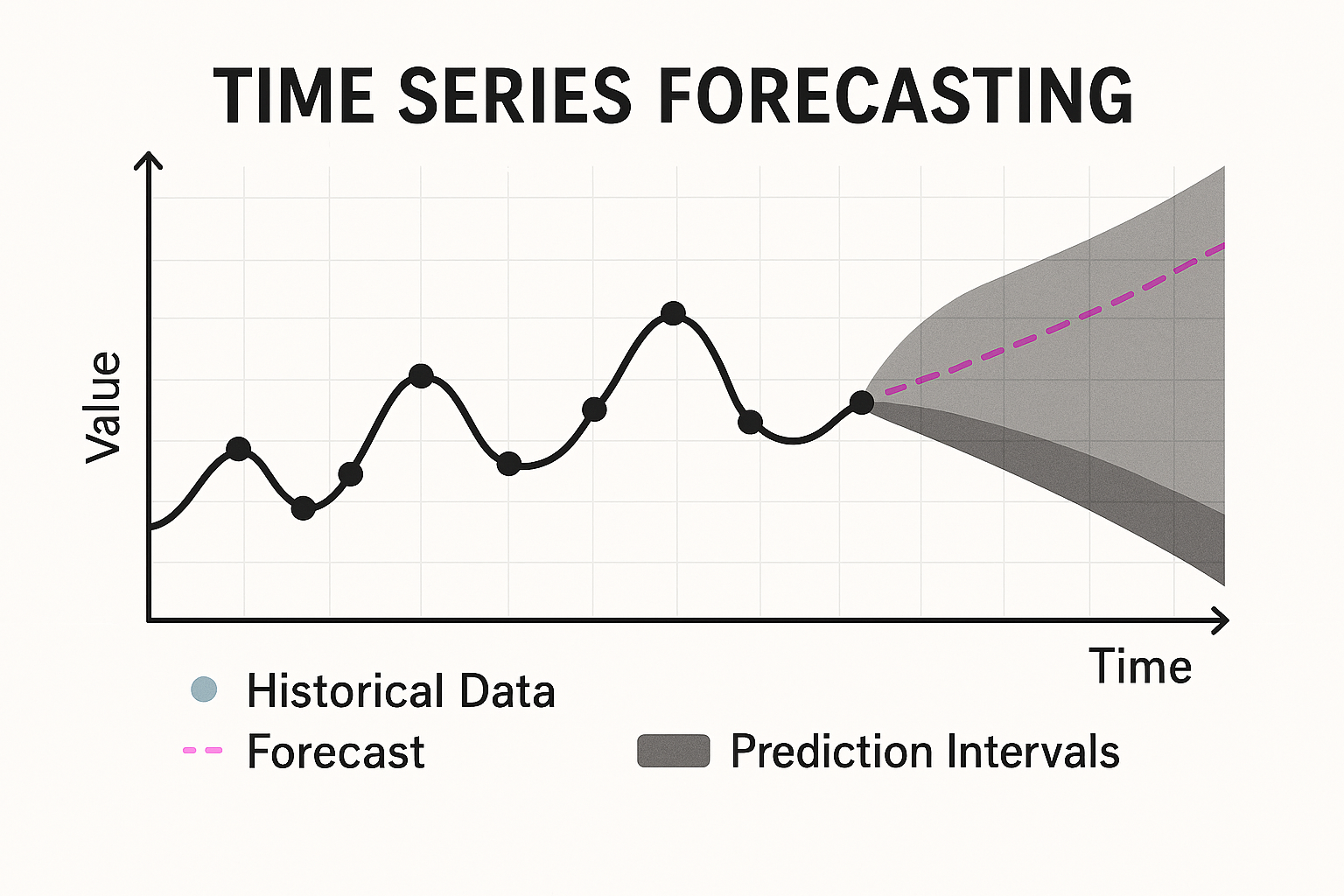}
    \caption{A conceptual visualization of time-series forecasting.}
    \label{fig:forecast}
\end{figure}

We considered one-dimensional time series with equally spaced time steps, where the task was to predict the next values based on a look-back window of the last $n$ time steps. As a starting point, we predicted simple functions such as sine waves or repeating values. For benchmarking stock data from Apple and Sales data from a pasta brand were used.

Many different models were compared. On the classical side, an LSTM, ARIMA and a basic method that just takes the last value as a prediction for the next value. As quantum models, three versions of a quantum neural network, a quantum deep Boltzmann machine (QDBM), quantum reservoir computing and a quantum LSTM were developed. All quantum models were run on a simulator several times to get stable results. To make this possible, Quriosity, BASF’s high-performance computing cluster, was used to handle this resource-intensive task. 

On Apple stock data, the simple last value prediction was the best model across all quantum and classical models, suggesting that this time series is essentially a random walk. On pasta sales data, the LSTM has the best performance, followed by the QNN.

The QDBM had large variations on different hyperparameter and on weight initialization. This can be seen as a weakness of the model or an opportunity to find better settings for future work. Also for other models, hyperparameter optimization had often a greater impact than the difference between the models.

\paragraph{Traffic-Light-based Evaluation}
\begin{itemize}
    \item \textbf{Green Light}
    \begin{itemize}
        \item \textbf{Scalability}: Quantum models need one qubit for each timestep in the look-back window. Thus already full experiments could run on today’s hardware: Quantum deep Boltzmann machines can be run on D-Wave’s quantum annealers, which enables already larger networks, up to several thousand qubits on real hardware today.% Larger look-back windows can lead to better performance but are not necessary to run a model.
        \item \textbf{Runtime \& Model and Implementation}: Quantum reservoir computing does not need training. After running a given amount of samples, the fitting process happens classically, which makes it highly suitable for early NISQ devices.
        \item Hybrid Methods: Hybrid networks with a quantum and a classical part can be implemented easily to gain advantages from both worlds in one model.
    \end{itemize}
    \item \textbf{Yellow Light}
    \begin{itemize}
        \item \textbf{Solution Quality}: The quantum models could achieve a comparable performance to the classical models. However, there is currently no clear sign of a possible advantage for any quantum model compared to a classical one.
        \item Dimensionality: Most real-world datasets have multiple dimensions, which cannot be evaluated on real hardware with a sufficiently large look-back window today. 
    \end{itemize}
    \item \textbf{Red Light}
    \begin{itemize}
        \item \textbf{Transferability}: Stock data is close to being a random walk, which means it cannot be predicted. This theory is supported by the fact, that the last-value method outperformed all models on this dataset, including the classical models, during our evaluation. Thus achieving results similar to classical models does not provide meaningful insights.
        %\item On the past dataset, the classical LSTM has the best overall performance
    \end{itemize}
\end{itemize}

\paragraph{Verdict}

\begin{itemize}
    \item \textbf{Current Viability}: The currently available hardware is sufficient for this simple example of a one-dimensional timeseries with a small look-back window. However, hardware access is costly, making full training with hundreds of epochs impractical. Most real-world applications would involve multiple dimensions, which are difficult to fit on real hardware.
    \item \textbf{Future Potential}: The future potential is high as this use case has many applications in different areas. At the same time machine learning is a black box and it is difficult to find general theoretical advantages.
    \item \textbf{Recommendation}: Keep in mind, that quantum machine learning is a viable tool for time series forecasting in the future. It can be seen as a new tool in the toolset of every machine learning expert and it is already useful for datasets with one dimension and when a small lookback window is sufficient. In 5 to 10 years it can be advanced to more complex examples.
\end{itemize}

\subsubsection{Multi-criteria Optimization}

Multi-criteria optimization aims to find the best combination of different input features to achieve desired values for the output features. This can be visualized as a Pareto front, where we get a multidimensional surface displaying the best input combinations.

\begin{figure}[ht]
    \centering
    \includegraphics[width=0.8\linewidth]{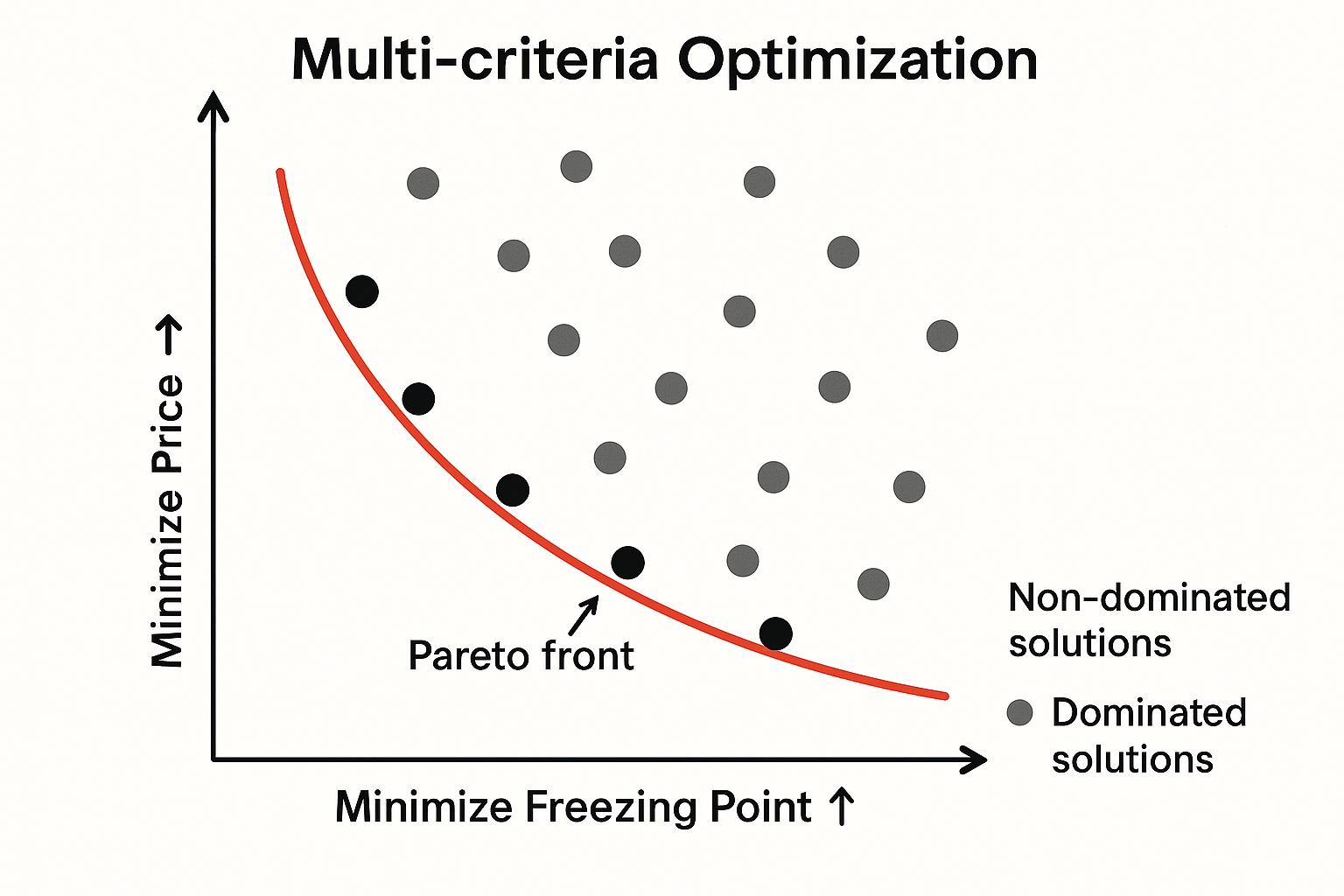}
    \caption{A conceptual visualization of multi-criteria optimization.}
    \label{fig:multicritopt}
\end{figure}

In this use case, we investigated a machine learning surrogate model designed to approximate the underlying relationship between inputs and outputs. This is very helpful if the experiments to get one datapoint are very expensive and fewer experiments are needed to get a good approximation of the efficient frontier.

The machine learning model used was a standard regression model in supervised learning. For classical comparison, random forest and a multi-layer perceptron are used. In addition, we built a quantum deep neural network composed of a feature map and an ansatz layer, forming a full layer. Data reuploading and trainable feature maps increased the expressive power of the model.

We tested the model on a synthetic dataset consisting of the functions \textit{perm}, \textit{rosenbrock}, \textit{rastrigin}, and \textit{schwefel}, which were combined into a multi-output function, requiring 4 qubits. For real-world applications we used a paint dataset provided by BASF, requiring 15 qubits for the 15 input features.

\paragraph{Traffic-Light-based Evaluation}

\begin{itemize}
    \item \textbf{Green Light}
    \begin{itemize}
        \item \textbf{Scalability}: The quantum model was trained on a simulator, and predictions were made on IonQ’s quantum gate-based computer. The results on real hardware were similar to the ones on the simulator when few layers were used. This suggests that larger models could also be trained and run once hardware becomes more affordable and less noisy.
        % \item The theory around quantum machine learning is quite new, but there are already many approaches that can improve the solution quality a lot.
        \item \textbf{Transferability}: This use case is a standard supervised learning task, namely regression via surrogate modeling to approximate an unknown function and predict new data points for multi-criterial optimization. As regression is among the most common ML tasks, the approach is widely transferable, though in quantum machine learning very high-dimensional inputs may exceed qubit resources and limit applicability.
    \end{itemize}
    \item \textbf{Yellow Light}
    \begin{itemize}
        \item \textbf{Solution Quality}: The quantum model achieved results comparable to the classical machine learning models. While the classical models performed slightly better, the difference was not substantial. It should be noted, however, that the classical baseline models were not fine-tuned, whereas the quantum models underwent extensive tuning. Thus, the performance of the classical models may be somewhat underestimated in this comparison.
        \item \textbf{Model and Implementation}: Barren plateaus are a complex issue similar to vanishing gradient problems in classical machine learning. There is no solution for it now and it might limit quantum machine learning, especially as this problem increases with larger circuits.
        \item \textbf{Runtime}: Training on a simulator is very slow and only small circuits are feasible right now. Real hardware is very expensive and a full training cycle of several thousand epochs would cost around 1 million euros right now on ion-based quantum computers.
    \end{itemize}
    %\\item \textbf{Red Light}
    %\\begin{itemize}
        %\item So far, there is no evidence that quantum machine learning offers an advantage over classical methods. We can test only small examples today, and it remains for further research how well it will work on larger and more complex instances.
    %\\end{itemize}
\end{itemize}

\paragraph{Verdict}

\begin{itemize}
    \item \textbf{Current Viability}: There is already hardware available to test specific use cases like the paint data on real hardware. However, limited reliability and uptime of current hardware remain significant barriers for consistent practical use. Future improvements in noise reduction would enable the use of deeper circuits and more qubits for complex tasks.
    \item \textbf{Future Potential}: This use case is highly relevant, as it is a common regression task with supervised learning and covers most machine learning tasks out there today. As hardware matures and larger datasets like images and text can be processed by quantum machine learning it has great potential to show advantages over classical machine learning at least in some areas. For industry, this could translate into substantial impact on product design.
    \item \textbf{Recommendation}: The most important task now is to investigate more into different ways how to set up the circuit and how to use machine learning techniques to prevent barren plateaus and to avoid local optima. It can already be used for datasets with few datapoints and low dimensions. As soon as the hardware matures it can be very interesting for a wide range of machine learning tasks. In parallel, careful benchmarking against well-tuned classical baselines will be essential to identify genuine advantages.
\end{itemize}

\subsection{Evaluation Overview of Considered Use Cases}
\label{sec:overview}

The industry applications evaluated during the project can broadly be categorized into three categories. An overview of the evaluation of all use cases is presented in Table 1.

The use cases best aligned with current quantum hardware and algorithms show comparable results to classical approaches for small test problem instances. For these, existing algorithms could already provide benefits compared to or in synergy with classical algorithms.

\textit{Production assignment and scheduling} is promising and practically viable in hybrid form, though quantum benefit is conditional on modeling and hardware, with limitations arising from the need to reformulate into binary quadratic form. Industrial relevance for instances with hundreds of thousand of qubits is already in the vicinity of hybrid solvers and estimated for around 2030 for pure quantum approaches according to roadmaps for quantum annealers and gate based devices.

\textit{Forecasting} of time series is feasible and shows comparable solution qualities to classical machine learning models in simple one-dimensional setups and has potential for more complex cases. However, it currently shows no clear advantage and lacks demonstrated scalability to multidimensional data. It is estimated that it could show industrial relevance within five to ten years once quantum hardware facilitates deeper circuits for tens to hundreds of logical qubits.

The \textit{cart pole quantum twin} is conceptually strong with broad transferability, though its scalability beyond toy models remains unproven and current implementations are much slower than classical ones. Similar to the forecasting use case, hardware needs to mature so that the number of layers can be increased for tens to hundreds of qubits.

\textit{Auto-carrier loading} is not yet competitive with classical heuristics, limited by inferior runtime and solution quality despite favorable scaling characteristics. For larger instances (e.g., fleet-level optimization) it could show benefits with larger and less error-prone future hardware. Industrial relevance could arise around 2030 according to current roadmaps judging on the requirement of hundreds of thousands of qubits for industrial relevant
problems.

\textit{Multi-criteria optimization} is viable for small datasets but limited by barren plateaus and heavy fine-tuning, which render scalability uncertain. Given hardware matures to support more layers which can be trained without the problem of barren plateaus we could envision a relevance for industrial problems in the early 2030s.

Overall, these use cases are limited less by modeling or algorithmic gaps and more by current qubit counts, noise, and circuit depth, with industrial relevance expected around or after 2030, ranging from early applicability in hybrid approaches to the need for deeper, less error-prone circuits.

The second category also features quantum algorithms that can tackle the full problem in a sufficient manner. However, it can already be estimated that for industrial relevant problem sizes the complexity increases strongly, rendering quantum approaches with the current technology infeasible. 

\textit{Modular production logistics} shows conditional promise for hybrid solvers, but the advantage is highly dependent on specific formulations while classical MILPs remain stronger for generic problems. Quantum hardware after 2030 could support solving industrial relevant problems which would require hundreds of thousand of qubits but current experiments show lower solution quality compared to classical solvers.

In the \textit{sensor positioning} use case, quantum methods are currently inferior to classical methods, mainly due to inflated qubit requirements from slack variables and poor stability of solutions. While the problem fits naturally into QUBO form, modeling inefficiencies severely limits scalability, and the solution quality lags behind classical solvers. This use case could become attractive for large-scale sensor grids only after 2035, as current roadmaps do not predict qubit counts and improved error correction to handle the inflated problem sizes until then.

Similarly, the \textit{cooling system optimization} use case is interesting for linearizable problems only. Such problems, however, are ample in the engineering domain. The quantum approach is inferior to classical alternatives until general systems (represented by the matrix and vector of the linear system) can be encoded efficiently. While runtime and scaling advantages for general cases remain purely theoretical under certain constraints (e.g., if all possible solutions need to be evaluated) the developed algorithm could present a runtime benefit which needs to be proven on future hardware that supports running the full algorithm for small instances which we estimate to be available not before 2035.

Finally, three use cases are challenged by insufficient modeling to facilitate the solution of full industrial problems on quantum hardware. The \textit{contextual bandit} has high theoretical impact but depends critically on solving open challenges in encoding and circuit efficiency; its limitation lies in the absence of general, efficient encodings for realistic data and simulations that are far slower than classical methods. Industrial relevance is unclear due to the open challenges and thus contingent on scalable encodings.

\textit{Train routing} shows promise through decomposition, but current models omit key real-world constraints and large instances remain intractable, making it clear that more effective modeling and algorithmic strategies are required. Similar to the contextual bandit, without being able to model the full set of constraints the industrial relevance remains unclear.

\textit{Pod coordination} is currently impractical due to the absence of modeling approaches that can incorporate continuous variables and the extreme qubit demand, exceeding 10,000 even for toy cases. These use cases face primary barriers in algorithmic innovation as well as hardware availability, with industrial relevance unclear or far-distant. Future developments in quantum algorithms and modeling will show if these use cases can benefit of quantum computing in the future.\\

\renewcommand{\arraystretch}{1.5} % increase row spacing
\newpage
\begin{tabular}{lccccc}
\toprule
\makecell{\textbf{Use}\\\textbf{Case}} 
& \makecell{Run-\\time} 
& \makecell{Scala-\\bility} 
& \makecell{Model \&\\Impl.} 
& \makecell{Solution\\Quality} 
& \makecell{Transfer-\\ability} \\
\midrule
\makecell{Forecast-\\ing} & \textcolor{Green}{Green} & \textcolor{Green}{Green} & \textcolor{Green}{Green} & \textcolor{amber}{Yellow} & \textcolor{purple}{Red} \\
\addlinespace[4pt]
\makecell{Production\\Assignment \&\\Scheduling} & \textcolor{Green}{Green} & \textcolor{Green}{Green} & \textcolor{amber}{Yellow} & \textcolor{amber}{Yellow} & \textcolor{purple}{Red} \\
\addlinespace[4pt]
\makecell{Cart Pole\\Quantum Twin} & \textcolor{purple}{Red} & \textcolor{amber}{Yellow} & \textcolor{amber}{Yellow} & \textcolor{Green}{Green} & \textcolor{Green}{Green} \\
\addlinespace[4pt]
\makecell{Auto-Carrier\\Loading } & \textcolor{amber}{Yellow} & \textcolor{amber}{Yellow} & \textcolor{Green}{Green} & \textcolor{amber}{Yellow} & \textcolor{amber}{Yellow} \\
\addlinespace[4pt]
\makecell{Multi-criteria\\Optimization} & \textcolor{amber}{Yellow} & \textcolor{Green}{Green} & \textcolor{amber}{Yellow} & \textcolor{amber}{Yellow} & \textcolor{Green}{Green} \\
\midrule
\makecell{Modular\\Production\\Logistics} & \textcolor{purple}{Red} & \textcolor{amber}{Yellow} & \textcolor{amber}{Yellow} & \textcolor{amber}{Yellow} & \textcolor{purple}{Red} \\
\addlinespace[4pt]
\makecell{Sensor\\Positioning} & \textcolor{amber}{Yellow} & \textcolor{purple}{Red} & \textcolor{amber}{Yellow} & \textcolor{purple}{Red} & \textcolor{Green}{Green} \\
\addlinespace[4pt]
\makecell{Cooling\\System\\Optimization} & \textcolor{purple}{Red} & \textcolor{purple}{Red} & \textcolor{purple}{Red} & \textcolor{amber}{Yellow} & \textcolor{Green}{Green} \\
\midrule
\makecell{Contextual\\Bandit} & \textcolor{purple}{Red} & \textcolor{purple}{Red} & \textcolor{purple}{Red} & \textcolor{amber}{Yellow} & \textcolor{Green}{Green} \\
\addlinespace[4pt]
\makecell{Train\\Routing} & \textcolor{purple}{Red} & \textcolor{amber}{Yellow} & \textcolor{purple}{Red} & \textcolor{purple}{Red} & \textcolor{amber}{Yellow} \\
\addlinespace[4pt]
\makecell{Pod\\Coordination} & \textcolor{purple}{Red} & \textcolor{purple}{Red} & \textcolor{amber}{Yellow} & \textcolor{purple}{Red} & \textcolor{purple}{Red} \\
\bottomrule

\end{tabular}
\vspace{1em}
\newline
\textbf{Table 1}: Overview of use cases evaluated in this work, separated in three categories from more to less optimistic about current and future quantum computing performance compared to classical methods.

\section{Recommendations for the Industry and\\ Stakeholders of the Quantum Ecosystem}
\label{sec:outlook}

Based on the use-case analyses and insights from this study, several recommendations can be derived for industry stakeholders, quantum technology providers, and the broader ecosystem. These recommendations span \mbox{short-,} mid-, and long-term perspectives, highlighting both practical actions and strategic directions for collaboration and innovation.

\begin{enumerate}
    \item In the short term, no clear benefits from pure quantum devices should be expected. The utility of hybrid devices should be investigated and thoroughly benchmarked against purely classical solutions, especially for optimization use cases that can be naturally mapped to QUBO formulations and do not suffer from significant overhead due to the requirement of slack variables. The evaluated use cases -- production assignment and scheduling, modular production logistics, and auto-carrier loading -- can serve as good reference problems.
    \item Quantum machine learning use cases that show results comparable to classical machine learning approaches should be continuously tested on improved hardware to determine if they scale similarly to their classical counterparts. Runtime and cost of solution remain important factors that require consideration and improvement. Joint investigations with hardware providers to optimize these metrics are recommended as valuable efforts. Reference examples from the investigated use case list include forecasting, the cart pole quantum twin, and multi-criteria optimization.
    \item Benefits for use cases in the second category depend on improvements in modeling and more efficient encoding. Research in these areas should be industrially motivated, with collaborations between industry, aca\-demia, algorithm developers, and hardware providers to develop optimized mappings and algorithms tailored to specific industrial use cases and hardware modalities. Several investigated use cases in this project fall into this category. The sensor positioning problem requires a more efficient encoding or one that does not rely on slack variables. The cooling system optimization use case would benefit from a more scalable system encoding. Additionally, the auto-carrier loading use case could benefit from more optimal modeling to achieve better solution quality and shorter runtimes. Specific hardware modalities combined with optimized modeling strategies could further improve the results obtained.
    \item The insights gained and implementations conducted in this project should serve as a blueprint for identifying additional industrial problems suitable for quantum computing. Building on these examples, both acade\-mia and industry can collaborate to develop reusable components, open-source frameworks, and standardized benchmarking protocols. These efforts will enable consistent tracking of progress and ensure comparability across different approaches. Collaborative initiatives such as QCHALLenge and QUTAC highlight the importance of aligning problem definitions, hardware roadmaps, and software toolchains among stakeholders. Continued cooperation will strengthen the European quantum ecosystem, foster reproducible experimentation, and accelerate the realization of industrially relevant quantum utility.
    \item Looking ahead, the evolution of quantum computing hardware is characterized by rapid advancements accompanied by ongoing uncertainties. Superconducting and gate-based architectures continue to improve through higher fidelities and modular designs, while quantum annealing remains valuable for heuristic exploration of combinatorial problems. Ion-trap systems offer high-fidelity operations, though their scalability to industrial levels remains open. Emerging approaches such as photonic, neuromorphic, and topological quantum computing show promise but currently have limited industrial integration. Over the next decade, progress is expected along three interdependent threads: solver-aware problem modeling that formulates industrial challenges compatible with both classical and quantum solvers; hardware maturation focused on embedding efficiency, error mitigation, and integration with classical control systems; and sustained collaboration among industry, academia, and hardware providers through shared benchmarks, data, and open frameworks like QUARK.
\end{enumerate}

\section*{Acknowledgments}

This paper was partially funded by the  German Federal Ministry of Research, Technology and Space (BMFTR) through the funding program “Quantum Computing – Applications for the Industry” (QCHALLenge) based on the allowance “Development of digital technologies” (contract number: 01MQ22008). AI tools (OpenAI ChatGPT) were used to support text refinement and visual material. The authors retained full responsibility for the scientific content.
\newpage
\printbibliography

@misc{qchallenge,
    title = {Projekt QCHALLenge, BMWK},
    url = {https://qarlab.de/qchallenge/},
    author = {QAR Lab},
    note = {[Accessed 15-07-2025]}
}

@misc{planqk,
    title = {PlanQK},
    url = {https://platform.planqk.de/community},
    author = {KIPU Quantum},
    note = {[Accessed 04-09-2025]}
}

@misc{luna,
    title = {Unlock Quantum-Powered Optimization with Luna},
    url = {https://aqarios.com/platform/},
    author = {Aqarios},
    note = {[Accessed 04-09-2025]}
}

@misc{qucun,
    title = {Das Quantum Computing User Network in Deutschland},
    url = {https://qucun.de},
    author = {QuCUN},
    note = {[Accessed 04-09-2025]}
}

@misc{railOscope,
    title = {Infrastracture},
    url = {https://railoscope.com/workspace/6074970aaba15e47fca2625d?capability=TcBEfyVmiJTWLZm8&branchId=6281fc03a6ce9f408c072307&noTT=true},
    author = {railOscope},
    note = {[Accessed 29-09-2025]}
}

@article{QUTAC,
	author = {{Quantum Technology and Application Consortium – QUTAC} and {Bayerstadler, Andreas} and {Becquin, Guillaume} and {Binder, Julia} and {Botter, Thierry} and {Ehm, Hans} and {Ehmer, Thomas} and {Erdmann, Marvin} and {Gaus, Norbert} and {Harbach, Philipp} and {Hess, Maximilian} and {Klepsch, Johannes} and {Leib, Martin} and {Luber, Sebastian} and {Luckow, Andre} and {Mansky, Maximilian} and {Mauerer, Wolfgang} and {Neukart, Florian} and {Niedermeier, Christoph} and {Palackal, Lilly} and {Pfeiffer, Ruben} and {Polenz, Carsten} and {Sepulveda, Johanna} and {Sievers, Tammo} and {Standen, Brian} and {Streif, Michael} and {Strohm, Thomas} and {Utschig-Utschig, Clemens} and {Volz, Daniel} and {Weiss, Horst} and {Winter, Fabian}},
	title = {Industry quantum computing applications},
	DOI= "10.1140/epjqt/s40507-021-00114-x",
	url= "https://doi.org/10.1140/epjqt/s40507-021-00114-x",
	journal = {EPJ Quantum Technol.},
	year = 2021,
	volume = 8,
	number = 1,
	pages = "25"
}

@misc{quark_github,
    title = {QUARK-framework},
    url = {https://github.com/QUARK-framework/QUARK-framework},
    author = {GitHub},
    note = {[Accessed 04-09-2025]}
}

@misc{qchallenge_github,
    title = {QCHALLenge-FW},
    url = {https://github.com/Project-QCHALLenge/QCHALLenge-FW},
    author = {GitHub},
    note = {[Accessed 29-09-2025]}
}

@misc{dwave_advantage,
	title = {{The AdvantageTM Quantum Computer}},
        author = {D-Wave},
	url = {https://www.dwavequantum.com/solutions-and-products/systems/},
        note = {[Accessed 15-07-2025]}
}

@misc{dwave,
	title = {{Quantum Realized}},
        author = {D-Wave},
	url = {https://www.dwavequantum.com},
        note = {[Accessed 15-07-2025]}
}

@misc{ibm_roadmap,
	author = {IBM Quantum Computing},
	title = {Technology for the Quantum Future},
	url ={https://www.ibm.com/quantum/technology#roadmap},
	note = {[Accessed 15-07-2025]}
}

@misc{iqm_roadmap,
    title = {Development Roadmap},
    url = {https://meetiqm.com/technology/roadmap/},
    author = {IQM},
    note = {[Accessed 15-07-2025]}
}

@misc{ionq_roadmap,
	author = {IonQ, Inc.},
	title = {{I}on{Q}'s {A}ccelerated {R}oadmap: {T}urning {Q}uantum {A}mbition into {R}eality --- ionq.com},
	howpublished = {\url{https://ionq.com/blog/ionqs-accelerated-roadmap-turning-quantum-ambition-into-reality}},
	note = {[Accessed 24-06-2025]},
}

@misc{quantinuum_roadmap,
	author = {Quantinuum},
	title = {{Q}uantinuum {U}nveils {A}ccelerated {R}oadmap to {A}chieve {U}niversal, {F}ully {F}ault-{T}olerant {Q}uantum {C}omputing by 2030},
	howpublished = {\url{https://www.quantinuum.com/press-releases/quantinuum-unveils-accelerated-roadmap-to-achieve-universal-fault-tolerant-quantum-computing-by-2030}},
	note = {[Accessed 24-06-2025]},
}

@misc{quantinuum_roadmap2,
	author = {Quantinuum},
	title = {Technical perspective: By the end of the decade, we will deliver universal, fully fault-tolerant quantum computing},
	howpublished = {\url{https://www.quantinuum.com/blog/technical-perspective-by-the-end-of-the-decade-we-will-deliver-universal-fault-tolerant-quantum-computing}},
	note = {[Accessed 15-08-2025]},
}

@misc{google_roadmap,
    title = {Our quantum computing roadmap},
    url = {https://quantumai.google/roadmap},
    author = {Google Quantum AI},
    note = {[Accessed 15-07-2025]}
}

@misc{sensor_paper,
      title={Quantum Annealing Hyperparameter Analysis for Optimal Sensor Placement in Production Environments}, 
      author={Nico Kraus and Marvin Erdmann and Alexander Kuzmany and Daniel Porawski and Jonas Stein},
      year={2025},
      eprint={2507.16584},
      archivePrefix={arXiv},
      primaryClass={cs.ET},
      url={https://arxiv.org/abs/2507.16584}, 
}

@article{Jaeck_ACL,
    author = {J{\"a}ck, Christian and G{\"o}nsch, Jochen and D{\"o}rmann-Osuna Hans},
    title = {Load Factor Optimization for the Auto Carrier Loading Problem},
    journal = {Transportation Science},
    volume = {57},
    number = {6},
    page = {1696-1719},
    year = {2023}
}

@misc{qilimanjaro,
      title={A scalable 2-local architecture for quantum annealing of Ising models with arbitrary dimensions}, 
      author={Ana Palacios and Artur Garcia-Saez and Bruno Julia-Diaz and Marta P. Estarellas},
      year={2025},
      eprint={2404.06861},
      archivePrefix={arXiv},
      primaryClass={quant-ph},
      doi={https://doi.org/10.1103/PhysRevApplied.23.054070},
      url={https://arxiv.org/abs/2404.06861}, 
}

@misc{qilimanjaro2,
    url = {https://www.qilimanjaro.tech/qilimanjaro-quantum-tech-and-do-it-now-selected-to-advance-european-quantum-computing-as-part-of-eurohpc-ju-initiative/},
    author = {Qilimanjaro},
    note = {Accessed August 5, 2025}
}

@misc{nec,
    title = {Accelerating Research with a Unique Approach NEC's Quantum Computer Research},
    url = {https://parityqc.com/a-new-quantum-annealer-by-nec},
    author = {ParityQC},
    note = {[Accessed 23-07-2025]},
}

@article{Harris_2010,
   title={Experimental demonstration of a robust and scalable flux qubit},
   volume={81},
   ISSN={1550-235X},
   url={http://dx.doi.org/10.1103/PhysRevB.81.134510},
   DOI={10.1103/physrevb.81.134510},
   number={13},
   journal={Physical Review B},
   publisher={American Physical Society (APS)},
   author={Harris, R. and Johansson, J. and Berkley, A. J. and Johnson, M. W. and Lanting, T. and Han, Siyuan and Bunyk, P. and Ladizinsky, E. and Oh, T. and Perminov, I. and Tolkacheva, E. and Uchaikin, S. and Chapple, E. M. and Enderud, C. and Rich, C. and Thom, M. and Wang, J. and Wilson, B. and Rose, G.},
   year={2010},
   month=apr }

@article{cplex,
  title={V12. 1: User’s Manual for CPLEX},
  author={Cplex, IBM ILOG},
  journal={International Business Machines Corporation},
  volume={46},
  number={53},
  pages={157},
  year={2009}
}

@Article{metrics2020009,
AUTHOR = {Soubra, Hassan and Elsayed, Hatem and Elbrolosy, Yousef and Adel, Youssef and Attia, Zeyad},
TITLE = {Comprehensive Review of Metrics and Measurements of Quantum Systems},
JOURNAL = {Metrics},
VOLUME = {2},
YEAR = {2025},
NUMBER = {2},
ARTICLE-NUMBER = {9},
URL = {https://www.mdpi.com/3042-5042/2/2/9},
ISSN = {3042-5042},
DOI = {10.3390/metrics2020009}
}

@misc{PsyQuantum,
      title={A manufacturable platform for photonic quantum computing}, 
      author={Koen Alexander and Andrea Bahgat and Avishai Benyamini and Dylan Black and Damien Bonneau and Stanley Burgos and Ben Burridge and Geoff Campbell and Gabriel Catalano and Alex Ceballos and Chia-Ming Chang and CJ Chung and Fariba Danesh and Tom Dauer and Michael Davis and Eric Dudley and Ping Er-Xuan and Josep Fargas and Alessandro Farsi and Colleen Fenrich and Jonathan Frazer and Masaya Fukami and Yogeeswaran Ganesan and Gary Gibson and Mercedes Gimeno-Segovia and Sebastian Goeldi and Patrick Goley and Ryan Haislmaier and Sami Halimi and Paul Hansen and Sam Hardy and Jason Horng and Matthew House and Hong Hu and Mehdi Jadidi and Henrik Johansson and Thomas Jones and Vimal Kamineni and Nicholas Kelez and Ravi Koustuban and George Kovall and Peter Krogen and Nikhil Kumar and Yong Liang and Nicholas LiCausi and Dan Llewellyn and Kimberly Lokovic and Michael Lovelady and Vitor Manfrinato and Ann Melnichuk and Mario Souza and Gabriel Mendoza and Brad Moores and Shaunak Mukherjee and Joseph Munns and Francois-Xavier Musalem and Faraz Najafi and Jeremy L. O'Brien and J. Elliott Ortmann and Sunil Pai and Bryan Park and Hsuan-Tung Peng and Nicholas Penthorn and Brennan Peterson and Matt Poush and Geoff J. Pryde and Tarun Ramprasad and Gareth Ray and Angelita Rodriguez and Brian Roxworthy and Terry Rudolph and Dylan J. Saunders and Pete Shadbolt and Deesha Shah and Hyungki Shin and Jake Smith and Ben Sohn and Young-Ik Sohn and Gyeongho Son and Chris Sparrow and Matteo Staffaroni and Camille Stavrakas and Vijay Sukumaran and Davide Tamborini and Mark G. Thompson and Khanh Tran and Mark Triplet and Maryann Tung and Alexey Vert and Mihai D. Vidrighin and Ilya Vorobeichik and Peter Weigel and Mathhew Wingert and Jamie Wooding and Xinran Zhou},
      year={2024},
      eprint={2404.17570},
      archivePrefix={arXiv},
      primaryClass={quant-ph},
      url={https://arxiv.org/abs/2404.17570}, 
}

@misc{Majorana,
      title={Roadmap to fault tolerant quantum computation using topological qubit arrays}, 
      author={David Aasen and Morteza Aghaee and Zulfi Alam and Mariusz Andrzejczuk and Andrey Antipov and Mikhail Astafev and Lukas Avilovas and Amin Barzegar and Bela Bauer and Jonathan Becker and Juan M. Bello-Rivas and Umesh Bhaskar and Alex Bocharov and Srini Boddapati and David Bohn and Jouri Bommer and Parsa Bonderson and Jan Borovsky and Leo Bourdet and Samuel Boutin and Tom Brown and Gary Campbell and Lucas Casparis and Srivatsa Chakravarthi and Rui Chao and Benjamin J. Chapman and Sohail Chatoor and Anna Wulff Christensen and Patrick Codd and William Cole and Paul Cooper and Fabiano Corsetti and Ajuan Cui and Wim van Dam and Tareq El Dandachi and Sahar Daraeizadeh and Adrian Dumitrascu and Andreas Ekefjärd and Saeed Fallahi and Luca Galletti and Geoff Gardner and Raghu Gatta and Haris Gavranovic and Michael Goulding and Deshan Govender and Flavio Griggio and Ruben Grigoryan and Sebastian Grijalva and Sergei Gronin and Jan Gukelberger and Jeongwan Haah and Marzie Hamdast and Esben Bork Hansen and Matthew Hastings and Sebastian Heedt and Samantha Ho and Justin Hogaboam and Laurens Holgaard and Kevin Van Hoogdalem and Jinnapat Indrapiromkul and Henrik Ingerslev and Lovro Ivancevic and Sarah Jablonski and Thomas Jensen and Jaspreet Jhoja and Jeffrey Jones and Kostya Kalashnikov and Ray Kallaher and Rachpon Kalra and Farhad Karimi and Torsten Karzig and Seth Kimes and Vadym Kliuchnikov and Maren Elisabeth Kloster and Christina Knapp and Derek Knee and Jonne Koski and Pasi Kostamo and Jamie Kuesel and Brad Lackey and Tom Laeven and Jeffrey Lai and Gijs de Lange and Thorvald Larsen and Jason Lee and Kyunghoon Lee and Grant Leum and Kongyi Li and Tyler Lindemann and Marijn Lucas and Roman Lutchyn and Morten Hannibal Madsen and Nash Madulid and Michael Manfra and Signe Brynold Markussen and Esteban Martinez and Marco Mattila and Jake Mattinson and Robert McNeil and Antonio Rodolph Mei and Ryan V. Mishmash and Gopakumar Mohandas and Christian Mollgaard and Michiel de Moor and Trevor Morgan and George Moussa and Anirudh Narla and Chetan Nayak and Jens Hedegaard Nielsen and William Hvidtfelt Padkær Nielsen and Frédéric Nolet and Mike Nystrom and Eoin O'Farrell and Keita Otani and Adam Paetznick and Camille Papon and Andres Paz and Karl Petersson and Luca Petit and Dima Pikulin and Diego Olivier Fernandez Pons and Sam Quinn and Mohana Rajpalke and Alejandro Alcaraz Ramirez and Katrine Rasmussen and David Razmadze and Ben Reichardt and Yuan Ren and Ken Reneris and Roy Riccomini and Ivan Sadovskyy and Lauri Sainiemi and Juan Carlos Estrada Saldaña and Irene Sanlorenzo and Simon Schaal and Emma Schmidgall and Cristina Sfiligoj and Marcus P. da Silva and Shilpi Singh and Sarat Sinha and Mathias Soeken and Patrick Sohr and Tomas Stankevic and Lieuwe Stek and Patrick Strøm-Hansen and Eric Stuppard and Aarthi Sundaram and Henri Suominen and Judith Suter and Satoshi Suzuki and Krysta Svore and Sam Teicher and Nivetha Thiyagarajah and Raj Tholapi and Mason Thomas and Dennis Tom and Emily Toomey and Josh Tracy and Matthias Troyer and Michelle Turley and Matthew D. Turner and Shivendra Upadhyay and Ivan Urban and Alexander Vaschillo and Dmitrii Viazmitinov and Dominik Vogel and Zhenghan Wang and John Watson and Alex Webster and Joseph Weston and Timothy Williamson and Georg W. Winkler and David J. van Woerkom and Brian Paquelet Wütz and Chung Kai Yang and Richard Yu and Emrah Yucelen and Jesús Herranz Zamorano and Roland Zeisel and Guoji Zheng and Justin Zilke and Andrew Zimmerman},
      year={2025},
      eprint={2502.12252},
      archivePrefix={arXiv},
      primaryClass={quant-ph},
      url={https://arxiv.org/abs/2502.12252}, 
}

@article{NAQC,
   title={Neutral atom quantum computing hardware: performance and end-user perspective},
   volume={10},
   ISSN={2196-0763},
   url={http://dx.doi.org/10.1140/epjqt/s40507-023-00190-1},
   DOI={10.1140/epjqt/s40507-023-00190-1},
   number={1},
   journal={EPJ Quantum Technology},
   publisher={Springer Science and Business Media LLC},
   author={Wintersperger, Karen and Dommert, Florian and Ehmer, Thomas and Hoursanov, Andrey and Klepsch, Johannes and Mauerer, Wolfgang and Reuber, Georg and Strohm, Thomas and Yin, Ming and Luber, Sebastian},
   year={2023},
   month=aug }

@misc{bitkom_talk1,
    url = {https://qarlab.de/en/talk-on-bitcom-2024},
    author = {QAR Lab},
    note = {Accessed September 30, 2025}
}

@misc{QCE_2024,
    url = {https://qarlab.de/en/qchallenge-publications-for-qce-conference-2024-in-montreal},
    author = {QAR Lab},
    note = {Accessed September 30, 2025}
}

@misc{Leap,
    url = {https://docs.dwavequantum.com/en/latest/industrial_optimization/leap_hybrid.html},
    author = {D-Wave},
    note = {Accessed October 27, 2025}
}

@misc{QCE_2025,
    url = {https://qarlab.de/en/contributions-to-qce-conference-2025},
    author = {QAR Lab},
    note = {Accessed September 30, 2025}
}

@misc{d-wave-advantage,
    url = {https://www.dwavequantum.com/media/3xvdipcn/14-1058a-a_advantage_processor_overview.pdf},
    author = {D-Wave},
    note = {Accessed August 8, 2025}
}

@misc{d-wave-advantage2,
    url = {https://www.dwavequantum.com/media/wakjcpsf/adv2_4400q_whitepaper-1.pdf},
    author = {D-Wave},
    note = {Accessed August 8, 2025}
}

@misc{d-wave-roadmap,
    url = {https://www.nextplatform.com/2025/03/31/d-wave-pushes-back-at-critics-shows-off-aggressive-quantum-roadmap/},
    author = {D-Wave},
    note = {Accessed August 8, 2025}
}

@misc{google_bristlecone,
    url = {https://research.google/blog/a-preview-of-bristlecone-googles-new-quantum-processor/},
    author = {Julian Kelly},
    note = {Accessed August 8, 2025}
}

@misc{ibm_eagle,
    url = {https://www.ibm.com/quantum/blog/127-qubit-quantum-processor-eagle},
    author = {Chow, Jerry and Dial, Oliver and Gambetta, Jay},
    note = {Accessed August 8, 2025}
}

@misc{ibm_condor,
    url = {https://www.ibm.com/quantum/blog/quantum-roadmap-2033},
    author = {Jay Gambetta},
    note = {Accessed August 8, 2025}
}

@misc{ibm_heron,
    url = {https://newsroom.ibm.com/2023-12-04-IBM-Debuts-Next-Generation-Quantum-Processor-IBM-Quantum-System-Two,-Extends-Roadmap-to-Advance-Era-of-Quantum-Utility},
    author = {IBM},
    note = {Accessed August 8, 2025}
}

@misc{ibm_heron2,
    url = {https://www.ibm.com/quantum/blog/qdc-2024},
    author = {Jay Gambetta and Ryan Mandelbaum},
    note = {Accessed August 8, 2025}
}

@misc{iqm_radiance,
    url = {https://meetiqm.com/products/iqm-radiance/},
    author = {IQM},
    note = {Accessed August 8, 2025}
}

@misc{ibm_nighthawk,
    url = {https://www.ibm.com/quantum/blog/large-scale-ftqc},
    author = {Ryan Mandelbaum and Jay Gambetta and Jerry Chow and Tushar Mittal and Theodore J. Yoder and Andrew Cross and Matthias Steffen},
    note = {Accessed August 8, 2025}
}

@article{google_9qubits,
   title={State preservation by repetitive error detection in a superconducting quantum circuit},
   volume={519},
   ISSN={1476-4687},
   url={http://dx.doi.org/10.1038/nature14270},
   DOI={10.1038/nature14270},
   number={7541},
   journal={Nature},
   publisher={Springer Science and Business Media LLC},
   author={Kelly, J. and Barends, R. and Fowler, A. G. and Megrant, A. and Jeffrey, E. and White, T. C. and Sank, D. and Mutus, J. Y. and Campbell, B. and Chen, Yu and Chen, Z. and Chiaro, B. and Dunsworth, A. and Hoi, I.-C. and Neill, C. and O’Malley, P. J. J. and Quintana, C. and Roushan, P. and Vainsencher, A. and Wenner, J. and Cleland, A. N. and Martinis, John M.},
   year={2015},
   month=mar, pages={66–69} }

@article{google_sycamore,
   title={Quantum supremacy using a programmable superconducting processor},
   volume={574},
   ISSN={1476-4687},
   url={http://dx.doi.org/10.1038/s41586-019-1666-5},
   DOI={10.1038/s41586-019-1666-5},
   number={7779},
   journal={Nature},
   publisher={Springer Science and Business Media LLC},
   author={Arute, Frank and Arya, Kunal and Babbush, Ryan and Bacon, Dave and Bardin, Joseph C. and Barends, Rami and Biswas, Rupak and Boixo, Sergio and Brandao, Fernando G. S. L. and Buell, David A. and Burkett, Brian and Chen, Yu and Chen, Zijun and Chiaro, Ben and Collins, Roberto and Courtney, William and Dunsworth, Andrew and Farhi, Edward and Foxen, Brooks and Fowler, Austin and Gidney, Craig and Giustina, Marissa and Graff, Rob and Guerin, Keith and Habegger, Steve and Harrigan, Matthew P. and Hartmann, Michael J. and Ho, Alan and Hoffmann, Markus and Huang, Trent and Humble, Travis S. and Isakov, Sergei V. and Jeffrey, Evan and Jiang, Zhang and Kafri, Dvir and Kechedzhi, Kostyantyn and Kelly, Julian and Klimov, Paul V. and Knysh, Sergey and Korotkov, Alexander and Kostritsa, Fedor and Landhuis, David and Lindmark, Mike and Lucero, Erik and Lyakh, Dmitry and Mandrà, Salvatore and McClean, Jarrod R. and McEwen, Matthew and Megrant, Anthony and Mi, Xiao and Michielsen, Kristel and Mohseni, Masoud and Mutus, Josh and Naaman, Ofer and Neeley, Matthew and Neill, Charles and Niu, Murphy Yuezhen and Ostby, Eric and Petukhov, Andre and Platt, John C. and Quintana, Chris and Rieffel, Eleanor G. and Roushan, Pedram and Rubin, Nicholas C. and Sank, Daniel and Satzinger, Kevin J. and Smelyanskiy, Vadim and Sung, Kevin J. and Trevithick, Matthew D. and Vainsencher, Amit and Villalonga, Benjamin and White, Theodore and Yao, Z. Jamie and Yeh, Ping and Zalcman, Adam and Neven, Hartmut and Martinis, John M.},
   year={2019},
   month=oct, pages={505–510} }

@article{google_willow,
   title={Quantum error correction below the surface code threshold},
   volume={638},
   ISSN={1476-4687},
   url={http://dx.doi.org/10.1038/s41586-024-08449-y},
   DOI={10.1038/s41586-024-08449-y},
   number={8052},
   journal={Nature},
   publisher={Springer Science and Business Media LLC},
   author={Acharya, Rajeev and Abanin, Dmitry A. and Aghababaie-Beni, Laleh and Aleiner, Igor and Andersen, Trond I. and Ansmann, Markus and Arute, Frank and Arya, Kunal and Asfaw, Abraham and Astrakhantsev, Nikita and Atalaya, Juan and Babbush, Ryan and Bacon, Dave and Ballard, Brian and Bardin, Joseph C. and Bausch, Johannes and Bengtsson, Andreas and Bilmes, Alexander and Blackwell, Sam and Boixo, Sergio and Bortoli, Gina and Bourassa, Alexandre and Bovaird, Jenna and Brill, Leon and Broughton, Michael and Browne, David A. and Buchea, Brett and Buckley, Bob B. and Buell, David A. and Burger, Tim and Burkett, Brian and Bushnell, Nicholas and Cabrera, Anthony and Campero, Juan and Chang, Hung-Shen and Chen, Yu and Chen, Zijun and Chiaro, Ben and Chik, Desmond and Chou, Charina and Claes, Jahan and Cleland, Agnetta Y. and Cogan, Josh and Collins, Roberto and Conner, Paul and Courtney, William and Crook, Alexander L. and Curtin, Ben and Das, Sayan and Davies, Alex and De Lorenzo, Laura and Debroy, Dripto M. and Demura, Sean and Devoret, Michel and Di Paolo, Agustin and Donohoe, Paul and Drozdov, Ilya and Dunsworth, Andrew and Earle, Clint and Edlich, Thomas and Eickbusch, Alec and Elbag, Aviv Moshe and Elzouka, Mahmoud and Erickson, Catherine and Faoro, Lara and Farhi, Edward and Ferreira, Vinicius S. and Burgos, Leslie Flores and Forati, Ebrahim and Fowler, Austin G. and Foxen, Brooks and Ganjam, Suhas and Garcia, Gonzalo and Gasca, Robert and Genois, Élie and Giang, William and Gidney, Craig and Gilboa, Dar and Gosula, Raja and Dau, Alejandro Grajales and Graumann, Dietrich and Greene, Alex and Gross, Jonathan A. and Habegger, Steve and Hall, John and Hamilton, Michael C. and Hansen, Monica and Harrigan, Matthew P. and Harrington, Sean D. and Heras, Francisco J. H. and Heslin, Stephen and Heu, Paula and Higgott, Oscar and Hill, Gordon and Hilton, Jeremy and Holland, George and Hong, Sabrina and Huang, Hsin-Yuan and Huff, Ashley and Huggins, William J. and Ioffe, Lev B. and Isakov, Sergei V. and Iveland, Justin and Jeffrey, Evan and Jiang, Zhang and Jones, Cody and Jordan, Stephen and Joshi, Chaitali and Juhas, Pavol and Kafri, Dvir and Kang, Hui and Karamlou, Amir H. and Kechedzhi, Kostyantyn and Kelly, Julian and Khaire, Trupti and Khattar, Tanuj and Khezri, Mostafa and Kim, Seon and Klimov, Paul V. and Klots, Andrey R. and Kobrin, Bryce and Kohli, Pushmeet and Korotkov, Alexander N. and Kostritsa, Fedor and Kothari, Robin and Kozlovskii, Borislav and Kreikebaum, John Mark and Kurilovich, Vladislav D. and Lacroix, Nathan and Landhuis, David and Lange-Dei, Tiano and Langley, Brandon W. and Laptev, Pavel and Lau, Kim-Ming and Le Guevel, Loïck and Ledford, Justin and Lee, Joonho and Lee, Kenny and Lensky, Yuri D. and Leon, Shannon and Lester, Brian J. and Li, Wing Yan and Li, Yin and Lill, Alexander T. and Liu, Wayne and Livingston, William P. and Locharla, Aditya and Lucero, Erik and Lundahl, Daniel and Lunt, Aaron and Madhuk, Sid and Malone, Fionn D. and Maloney, Ashley and Mandrà, Salvatore and Manyika, James and Martin, Leigh S. and Martin, Orion and Martin, Steven and Maxfield, Cameron and McClean, Jarrod R. and McEwen, Matt and Meeks, Seneca and Megrant, Anthony and Mi, Xiao and Miao, Kevin C. and Mieszala, Amanda and Molavi, Reza and Molina, Sebastian and Montazeri, Shirin and Morvan, Alexis and Movassagh, Ramis and Mruczkiewicz, Wojciech and Naaman, Ofer and Neeley, Matthew and Neill, Charles and Nersisyan, Ani and Neven, Hartmut and Newman, Michael and Ng, Jiun How and Nguyen, Anthony and Nguyen, Murray and Ni, Chia-Hung and Niu, Murphy Yuezhen and O’Brien, Thomas E. and Oliver, William D. and Opremcak, Alex and Ottosson, Kristoffer and Petukhov, Andre and Pizzuto, Alex and Platt, John and Potter, Rebecca and Pritchard, Orion and Pryadko, Leonid P. and Quintana, Chris and Ramachandran, Ganesh and Reagor, Matthew J. and Redding, John and Rhodes, David M. and Roberts, Gabrielle and Rosenberg, Eliott and Rosenfeld, Emma and Roushan, Pedram and Rubin, Nicholas C. and Saei, Negar and Sank, Daniel and Sankaragomathi, Kannan and Satzinger, Kevin J. and Schurkus, Henry F. and Schuster, Christopher and Senior, Andrew W. and Shearn, Michael J. and Shorter, Aaron and Shutty, Noah and Shvarts, Vladimir and Singh, Shraddha and Sivak, Volodymyr and Skruzny, Jindra and Small, Spencer and Smelyanskiy, Vadim and Smith, W. Clarke and Somma, Rolando D. and Springer, Sofia and Sterling, George and Strain, Doug and Suchard, Jordan and Szasz, Aaron and Sztein, Alex and Thor, Douglas and Torres, Alfredo and Torunbalci, M. Mert and Vaishnav, Abeer and Vargas, Justin and Vdovichev, Sergey and Vidal, Guifre and Villalonga, Benjamin and Heidweiller, Catherine Vollgraff and Waltman, Steven and Wang, Shannon X. and Ware, Brayden and Weber, Kate and Weidel, Travis and White, Theodore and Wong, Kristi and Woo, Bryan W. K. and Xing, Cheng and Yao, Z. Jamie and Yeh, Ping and Ying, Bicheng and Yoo, Juhwan and Yosri, Noureldin and Young, Grayson and Zalcman, Adam and Zhang, Yaxing and Zhu, Ningfeng and Zobrist, Nicholas},
   year={2024},
   month=dec, pages={920–926} }

@article{iqm_spark,
   title={On-premises superconducting quantum computer for education and research},
   volume={11},
   ISSN={2196-0763},
   url={http://dx.doi.org/10.1140/epjqt/s40507-024-00243-z},
   DOI={10.1140/epjqt/s40507-024-00243-z},
   number={1},
   journal={EPJ Quantum Technology},
   publisher={Springer Science and Business Media LLC},
   author={Rönkkö, Jami and Ahonen, Olli and Bergholm, Ville and Calzona, Alessio and Geresdi, Attila and Heimonen, Hermanni and Heinsoo, Johannes and Milchakov, Vladimir and Pogorzalek, Stefan and Sarsby, Matthew and Savytskyi, Mykhailo and Seegerer, Stefan and Šimkovic, Fedor and Sriluckshmy, P. V. and Vesanen, Panu T. and Nakahara, Mikio},
   year={2024},
   month=apr }

@INPROCEEDINGS{PAS,
  author={Awasthi, Abhishek and Kraus, Nico and Krellner, Florian and Zambrano, David},
  booktitle={2024 IEEE International Conference on Quantum Computing and Engineering (QCE)}, 
  title={Real World Application of Quantum-Classical Optimization for Production Scheduling}, 
  year={2024},
  volume={02},
  number={},
  pages={239-244}
}

@inproceedings{moprolog,
    author = "Krellner, Florian and Awasthi, Abhishek and Kraus, Nico and Braun, Sarah and Poppel, Michael and Porawski, Daniel",
    title = "{Solving a real-world modular logistic scheduling problem with a quantum-classical metaheuristics}",
    booktitle = "{2025 International Conference on Quantum Computing and Engineering (QCE)}",
    eprint = "2507.21701",
    archivePrefix = "pre-print - arXiv",
    primaryClass = "quant-ph",
    month = "7",
    year = "2025"
}

@misc{hölscher2025cooling,
      title={Quantum Simulation-Based Optimization of a Cooling System}, 
      author={Leonhard Hölscher and Lukas Müller and Or Samimi and Tamuz Danzig},
      year={2025},
      eprint={2504.15460},
      archivePrefix={arXiv},
      primaryClass={quant-ph},
      url={https://arxiv.org/abs/2504.15460}, 
}

@inproceedings{Hein_2017,
   title={A benchmark environment motivated by industrial control problems},
   url={http://dx.doi.org/10.1109/SSCI.2017.8280935},
   DOI={10.1109/ssci.2017.8280935},
   booktitle={2017 IEEE Symposium Series on Computational Intelligence (SSCI)},
   publisher={IEEE},
   author={Hein, Daniel and Depeweg, Stefan and Tokic, Michel and Udluft, Steffen and Hentschel, Alexander and Runkler, Thomas A. and Sterzing, Volkmar},
   year={2017},
   month=nov, pages={1–8} }

@misc{sun2025,
      title={First Experience with Real-Time Control Using Simulated VQC-Based Quantum Policies}, 
      author={Yize Sun and Mohamad Hagog and Marc Weber and Daniel Hein and Steffen Udluft and Volker Tresp and Yunpu Ma},
      year={2025},
      eprint={2508.01690},
      archivePrefix={arXiv},
      primaryClass={quant-ph},
      url={https://arxiv.org/abs/2508.01690}, 
}

@misc{Schulte2025,
    title = {Variational Quantum Circuits in Offline Contextual Bandit Problems},
    author = {Lukas Schulte and Daniel Hein and Steffen Udluft and Thomas A. Runkler},
    year = {2025},
    eprint={2509.07633},
    archivePrefix={arXiv},
    primaryClass={quant-ph},
    url={https://arxiv.org/abs/2509.07633}
}

@article{Eder2025,
  title = {Efficient active–passive vehicle coordination in multimodal transportation networks},
  volume = {203},
  ISSN = {1366-5545},
  url = {http://dx.doi.org/10.1016/j.tre.2025.104318},
  DOI = {10.1016/j.tre.2025.104318},
  journal = {Transportation Research Part E: Logistics and Transportation Review},
  publisher = {Elsevier BV},
  author = {Eder,  Peter J. and Ramoser,  Simon and Braun,  Sarah and Weltge,  Stefan},
  year = {2025},
  month = nov,
  pages = {104318}
}
\end{document}